\documentclass[12pt]{iopart}

\expandafter\let\csname equation*\endcsname\relax
\expandafter\let\csname endequation*\endcsname\relax

\usepackage{amsmath}
\usepackage{graphicx}
\usepackage{amssymb}
\usepackage{xcolor}
\usepackage{multirow}
\usepackage[utf8x]{inputenc}
\usepackage{hyperref}

\newcommand{\eHL}{\epsilon_{HL}}    
\newcommand{\pol}{\langle \tilde{\alpha} \rangle }
\newcommand{\MM}{{\cal M} }
\newcommand{\btheta}{\boldsymbol{\theta}}
\newcommand{\bb}{\mathbf{b}}
\newcommand{\bub}{\mathbf{\underbar{b}}}
\newcommand{\BFGS}{\texttt{BFGS}}

\let\oldAA\AA
\renewcommand{\AA}{\text{\normalfont\oldAA}}
\def\munderbar#1{\underline{\sbox\tw@{$#1$}\dp\tw@\z@\box\tw@}}

\begin{document}

\title[]{Inverse molecular design and parameter optimization with H\"uckel theory using automatic differentiation}

\author{Rodrigo A. Vargas--Hern\'andez}
\address{Chemical Physics Theory Group, Department of Chemistry, University of Toronto, Toronto, Ontario, M5S 3H6, Canada.}
\address{Vector Institute for Artificial Intelligence, 661 University Ave. Suite 710, Toronto, Ontario M5G 1M1, Canada.}
\address{current affiliation: Department of Chemistry and Chemical Biology, McMaster University, 1280 Main Street West, Hamilton, Ontario, L8S 4M1 Canada.}
\ead{vargashr@mcmaster.ca}

\author{Kjell Jorner}
\address{Chemical Physics Theory Group, Department of Chemistry, University of Toronto, Toronto, Ontario, M5S 3H6, Canada.}
\address{Department of Computer Science, University of Toronto, 40 St. George St, Ontario M5S 2E4, Canada.}
\address{Department of Chemistry and Chemical Engineering, Chalmers University of Technology, Kemigården 4, SE-41258, Gothenburg, Sweden.}

\author{Robert Pollice}
\address{Chemical Physics Theory Group, Department of Chemistry, University of Toronto, Toronto, Ontario, M5S 3H6, Canada.}
\address{Department of Computer Science, University of Toronto, 40 St. George St, Ontario M5S 2E4, Canada.}
\address{current affiliation: Stratingh Institute for Chemistry, University of Groningen, Nijenborgh 4, Groningen, 9747 AG, The Netherlands.}

\author{Al\'an Aspuru--Guzik}
\address{Chemical Physics Theory Group, Department of Chemistry, University of Toronto, Toronto, Ontario, M5S 3H6, Canada.}
\address{Department of Computer Science, University of Toronto, 40 St. George St, Ontario M5S 2E4, Canada.}
\address{Department of Chemical Engineering \& Applied Chemistry, 200 College St., University of Toronto, Ontario M5S 3E5, Canada.}
\address{Department of Materials Science \& Engineering, 184 College St., University of Toronto, Ontario M5S 3E4, Canada.}
\address{Vector Institute for Artificial Intelligence, 661 University Ave. Suite 710, Toronto, Ontario M5G 1M1, Canada.}
\address{Lebovic Fellow, Canadian Institute for Advanced Research (CIFAR), 661 University Ave., Toronto, Ontario M5G 1M1, Canada.}

\vspace{10pt}

\begin{abstract}
Semi-empirical quantum chemistry has recently seen a renaissance with applications in high-throughput virtual screening and machine learning. The simplest semi-empirical model still in widespread use in chemistry is Hückel's $\pi$-electron molecular orbital theory. In this work, we implemented a Hückel program using differentiable programming with the \texttt{JAX} framework, based on limited modifications of a pre-existing NumPy version. The auto-differentiable Hückel code enabled efficient gradient-based optimization of model parameters tuned for excitation energies and molecular polarizabilities, respectively, based on as few as 100 data points from density functional theory simulations. In particular, the facile computation of the polarizability, a second-order derivative, via auto-differentiation shows the potential of differentiable programming to bypass the need for numeric differentiation or derivation of analytical expressions. Finally, we employ gradient-based optimization of atom identity for inverse design of organic electronic materials with targeted orbital energy gaps and polarizabilities. Optimized structures are obtained after as little as 15 iterations, using standard gradient-based optimization algorithms.
\end{abstract}

%
%
%
%
%


\section{Introduction}
\label{sec:intro}

Mathematical models that are both predictive and provide insight are a cornerstone of the physical sciences. However, accurate models for complicated processes often have no analytical solution and require large  computational resources to solve numerically. At the same time, they also tend to be hard to interpret, as highlighted by Mulliken's famous quote "the more accurate the calculations became, the more the concepts tended to vanish into thin air" \cite{mulliken_1965}. Approximate models with problem-specific parameters are therefore used in practice, but finding optimal values for these parameters can be non-trivial. Parameter optimization normally requires considerable amounts of reference data and is done either manually or with algorithms that do not take advantage of first or higher order derivatives as the corresponding analytical expressions are often unavailable.

In chemistry, the Schrödinger equation is an archetype of such a mathematical model that describes the interactions between nuclei and electrons in both atoms and molecules. However, (near) exact solutions are too computationally expensive for most molecules of interest. Quantum chemistry is an entire research field dedicated to finding computationally efficient solutions to the Schrödinger equation by introducing prudent approximations or reformulations \cite{lowdin_quantum_1957}. One approach that was extremely successful in the early days of quantum chemistry is the use of so-called \textit{semiempirical} (SE) approximations \cite{semiqm:thiel:2014}. The central idea is the use of problem-specific parameters to simplify the mathematical form of the Schrödinger equation. One of the earliest SE models was Hückel's method to treat the $\pi$-electrons in organic molecules \cite{huckel:1931,huckel:1931_2,huckel:1932,huckel:1933}. Traditionally, the parameters in the Hückel method were derived manually by human scientists with the aim to reproduce properties for well-known reference molecules \cite{streitwieser_1961, hess_1972}, or they were derived from more accurate calculations \cite{van-catledgePariserParrPoplebasedSetHueckel1980}. Over the years, the Hückel method has been used for pedagogical purposes and for obtaining physical insight into problems in organic chemistry \cite{fleming_2010} and photochemistry and photophysics \cite{klan_2009, anstoter_2022}. However, it can also be used as a fast method for the prediction of molecular properties \cite{gomes_aromaticity_2001}, and for inverse design of molecules with desired target properties \cite{sanchez_inverse_2018, Beratan_jcp_2008}.

The recent upsurge in machine learning (ML), and specifically deep neural networks, created a need for robust and efficient algorithms to co-optimize a very large number of model parameters for various architectures. This problem is now solved by automatic differentiation (AD), a technique to evaluate the derivatives of mathematical expressions via the chain rule \cite{ad_survey}. Importantly, AD removes the need to determine analytic expressions for derivatives and makes complicated mathematical models amenable to gradient-based optimization, allowing them to be applied in the same way as general supervised machine learning models. Regular machine learning approaches like deep neural networks are meant to be very general mathematical models with a large number of parameters. Through learning, they can adapt to essentially any problem provided sufficient training data is available. In contrast, physics-based mathematical models have expressions that are specific to a certain type of problem to be solved and feature a much smaller number of parameters. Implementing physical models such as quantum chemistry within AD frameworks enables the use of default learning algorithms for parameter optimization with a potentially much smaller training data requirement. Along these lines, autodifferentiable versions of Hartree-Fock \cite{tamayo:2018,pyscfad}, density functional theory (DFT) \cite{dqc,AD_DFT_XC_NN,AD_DFT_XC_NN_2,pyscfad}, excited state mean-field theory \cite{AD_Neuscamman}. For  semi-empirical methods \cite{zhou_graphics_2020,huckel:nn:jcp,semiempirical:JPClett:2022,DL:semiempiricals:PNAS:2022}, and other applications \cite{ad:jcp:2015,dqc,quax,pennylane_qchem,pennylane_qchem, ad:ccd,ad:alchemy,ML-Chem_NatRevChem,Vargas:2021_AD,Vargas:2020_AD,Naruki_ADQC,AD_NucQC}, AD has been used to accelerate the calculation of gradients physical methods and to blend with ML algorithms.

In this work, we developed an auto-differentiable implementation of the Hückel method, by minimal adaptation of an initially developed NumPy \cite{numpy_2020} version into the \texttt{JAX} \cite{jax2018github} AD framework. We use this model to demonstrate the ease and efficiency of parameter fitting based on computational reference data sets for both excitation energies and molecular polarizabilities, a property calculated via a second order derivative. Additionally, we demonstrate that our AD model allows for gradient-based inverse design by regarding the atomic composition of a molecular system as an adjustable parameter to find molecules with targeted properties.\cite{Beratan_jcp_2008} The corresponding code is made publicly available, allowing it to be applied to a large variety of chemical problems. As the Hückel calculations are extremely fast, our workflow allows for facile development of property-specific models that can be readily used in molecular generative models that require on the order of 10\textsuperscript{5}--10\textsuperscript{6} property evaluations.

The paper is structured as follows: we first present a short introduction on automatic differentiation and the H\"uckel model (Sections \ref{sec:ad} and \ref{sec:huckel}). Following that, we execute inverse design of molecules as a fully differentiable procedure (Section \ref{sec:inv_design}) and perform optimization of the H\"uckel model parameters using modern gradient-based methods (Section \ref{sec:param_optimization}).


\section{Methods} \label{sec:methods}

\subsection{Automatic differentiation} \label{sec:ad}

Gradients and high-order derivatives are at the core of physical simulations. For physical models, common approaches to evaluate derivatives of any order are closed-form solutions, symbolic differentiation, and numerical differentiation, \textit{i.e.}, finite differences \cite{diffphys,ad_survey}. For any function represented as a computer program, AD \cite{ad_survey} is an alternative way to compute gradients and higher order derivatives. AD makes use of the chain rule for differentiation to create a program that computes the gradients during evaluation. There are two main modes in AD, \textit{forward} and \textit{reverse} mode. For scalar functions, reverse mode is more efficient as differentiation requires a single evaluation of the function to fully compute the Jacobian. An example of reverse mode differentiation is the backpropagation algorithm that is used for training neural networks. For more details about AD, we refer the reader to Ref. \cite{ad_survey}. 

The optimization of ML models is mostly done with methods that require the gradient of the \textit{loss} or \textit{error function} (${\cal L}$) with respect to the model parameters ($\boldsymbol{\theta}$), $\nabla_{\boldsymbol{\theta}} {\cal L}(\boldsymbol{\theta})$. All contemporary ML libraries, \textit{e.g.}, \texttt{Tensorflow} \cite{tensorflow:2015}, \texttt{PyTorch} \cite{pytorch} and \texttt{JAX} \cite{jax2018github}, are built on top of an AD engine which computes $\nabla_{\boldsymbol{\theta}} {\cal L}(\boldsymbol{\theta})$ for any ML model. Given the robustness of AD libraries, differentiating physical models \cite{diffphys} could be done similarly to modern ML algorithms. 

\subsection{H\"uckel model} \label{sec:huckel}

The H\"uckel model, a well-known semi-empirical quantum chemistry model \cite{huckel:1931,huckel:1931_2,huckel:1932,huckel:1933}, was first proposed to describe the interactions of $\pi$-electrons in conjugated unsaturated hydrocarbons. In the H\"uckel model, this interaction is restricted to electrons centered at nearest neighbour atoms. Generally, the H\"uckel model is considered a tight-binding type Hamiltonian (Eq. \ref{eqn:tb_ham}) where the on-site and hopping parameters are commonly denoted in the literature as $\alpha_{\ell}$ and $\beta_{\ell,k}$, respectively. The matrix elements of the H\"uckel Hamiltonian are given by

\begin{equation}
    H_{\ell,k} = \begin{cases}
 \alpha_{\ell}, &  \ell=k \\
 \beta_{\ell,k}, &  \ell \text{   and   } k \text{  are adjacent  } \\
 0, & \text{ otherwise, }
\end{cases}
\label{eqn:tb_ham}
\end{equation}
where the $\alpha_{\ell}$ parameters roughly represent the energy of an electron in a 2\texttt{p} orbital, and the $\beta_{\ell,k}$ parameters describe the energy of an electron in the bond $\ell-k$. Extensions of the H\"uckel Hamiltonian are possible and can, for instance, incorporate distance-dependence via $\beta_{\ell,k} = \beta^0_{\ell,k}\;g(\mathbf{R}_{\ell,k})$ (cf. Section \ref{sec:param_optimization}). For more details, we refer the reader to standard quantum chemistry textbooks \cite{yates:1978,mobook:pople1970}.

Notably, any molecular property computed with the H\"uckel method depends directly on the $\alpha_{\ell}$ and $\beta_{\ell,k}$ parameters. Therefore, by tuning their values, one can either construct a more accurate H\"uckel model for a given molecule and property (cf. Section \ref{sec:param_optimization}), or search for atomic compositions that optimize target properties given a preset connectivity (cf. Section \ref{sec:inv_design}). In the following sections, we demonstrate how AD can be used to facilitate both these types of problems.


\section{Results and Discussion} \label{sec:results}

\subsection{Inverse molecular design} \label{sec:inv_design}

\begin{figure}
    \centering
    \includegraphics[scale=1.0]{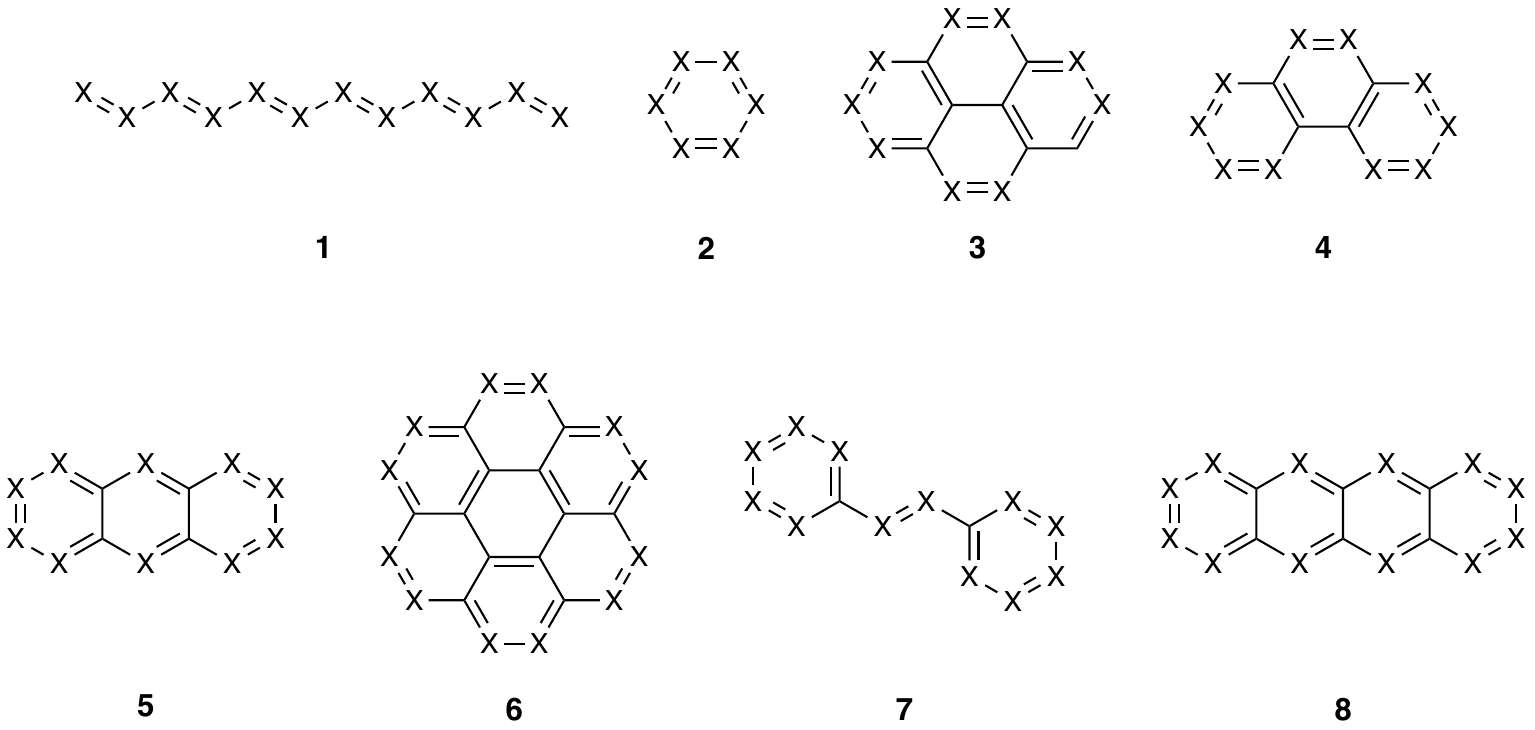}
    \caption{The eight molecular frameworks considered for inverse molecular design in this work \cite{Beratan_jcp_2008}. The $x$-symbol represents atomic sites whose identity is optimized.}
    \label{fig:molecules}
\end{figure}

Inverse molecular design can be carried out via gradient-based optimization methods, as shown in Ref. \cite{Beratan_jcp_2008}. The H\"uckel model can be extended to search for the molecular structure with a desired property. Both the diagonal and off-diagonal elements of the H\"uckel Hamiltonian matrix can be described by a weighted average of atom types at each site,
\begin{equation}
    H_{\ell,k} = \begin{cases}
 \sum_{i}^{M}b_{\ell}^{i}\alpha^i_{\ell}, &  \ell=k \\
 \sum_{i}^{M}\sum_{j}^{M}b_{\ell}^{i}b_{k}^{j}\beta^{ij}_{\ell,k}, &  \ell \text{   and   } k \text{  are adjacent  } \\
 0, & \text{ otherwise },
\end{cases}
\label{eqn:tb_ham_invdesing}
\end{equation}
where $b_{\ell}^{i}$ is the weight of the atom of type $i$ for site $\ell$. 
For a meaningful description, the weights of each site must be normalized, \textit{i.e.}, $\sum_{i}^{M}b_{\ell}^{i} = 1$. $M$ is the total number of atom types considered in the search. As a proof of concept, we consider eight different molecular frameworks \cite{Beratan_jcp_2008}, which are displayed in Fig. \ref{fig:molecules}. The $x$-symbol indicates the sites with variable atom types to be optimized. 
We only considered carbon ($C$), nitrogen ($N$) and phosphorus ($P$), i.e., $M=3$, as these atom types each contribute one electron, assuming that the remaining valences of carbon will be satisfied with a bond to an implicit hydrogen atom, and can be incorporated interchangeably at all sites with two neighbors in the $\pi$-framework (Fig. \ref{fig:molecules}). Therefore we defined the following vector of atom type weight parameters: $\bb_{\ell} = [b_{\ell}^{C},b_{\ell}^{N},b_{\ell}^{P}]$.
For clarity, $\bb$ jointly describes the $\bb_{\ell}$ parameters for all search sites in a molecule, \textit{i.e.}, $\bb=\{\bb_{\ell}\}^N$. 
For all results presented, the values of the $\alpha_\ell$ and $\beta^{0}_{\ell,k}$ parameters were previously optimized with respect to the desired property (cf. Section \ref{sec:param_optimization}). \\

For the set of eight molecules considered (\textit{cf.} Fig. \ref{fig:molecules}), we search for the type of atom $\bb$ at each site that gives the lowest HOMO-LUMO gap (Eq. \ref{eqn:homo_lumo}), denoted as $\epsilon_{HL}$, and the maximum polarizability denoted as $\pol$, (Eq, \ref{eqn:polarizability}).
$\epsilon_{HL}$ is defined as,
\begin{equation}
    \eHL = \epsilon_{LUMO} - \epsilon_{HOMO},
    \label{eqn:homo_lumo}
\end{equation}
where $\epsilon_{HOMO}$ and $\epsilon_{LUMO}$ are the eigenvalues of the highest occupied molecular orbital (HOMO), and lowest unoccupied molecular orbital (LUMO), respectively.
The polarizability function is defined as
\begin{equation}
    \pol = \frac{1}{3}\left (\tilde{\alpha}_{xx} + \tilde{\alpha}_{yy} + \tilde{\alpha}_{zz} \right),
    \label{eqn:polarizability}
\end{equation}
where the $\tilde{\alpha}_{ij}$ elements are the polarizability components defined as
\begin{equation}
    \tilde{\alpha}_{ij} = - \frac{\partial^2 E}{\partial F_{i} \partial F_{j}}.
    \label{eqn:polarizability_element}
\end{equation}
The $F_i$ terms are the components of the electric field, $\vec{\boldsymbol{F}} = [F_{x},F_{y},F_{z}]$, and $E$ is the electronic energy of the system. 
The elements of the polarizability tensor are usually computed using a finite-difference (FD) approach  \cite{Beratan_jcp_2008}, 
\begin{equation}
  \tilde{\alpha}_{ii} = \frac{2E(0)-\left [E(-F_{i})+ E(+F_{i})\right]}{F_{i}^2},
  \label{eqn:pol_fd}
\end{equation}
where the electronic energy is evaluated several times, typically three times for each diagonal element, and four times for each cross term.
Notably, if the parameters $\bb$ are to be optimized using a gradient-based method, the Jacobians $\nabla_{\bb}\eHL$ and $\nabla_{\bb}\pol$ will also be constructed using an FD approach. However, this will increase the number of energy evaluations needed, especially for $\pol$ as the elements of $\nabla_{\bb}\pol$ are third-order derivatives: $ \frac{\partial }{\partial \bb^{i}} \frac{\partial^2 \tilde{\alpha}_{kk}}{\partial F^2_{k} }$.
Thus, for a single element of $\nabla_{\bb}\pol$, using FD will require 18 energy calculations, ${\cal O}(18\times\| \bb \|)$, where $\| \bb \|$ is the total number of parameters in $\bb$. For $\eHL$, using FD, we only require ${\cal O}(2\times\| \bb \|)$, as $\nabla_{\bb}\eHL$ is a first-order derivative.

In contrast, using modern AD frameworks, we can efficiently compute, with a single energy calculation (\textit{i.e.}, one forward pass), the Jacobian of $\eHL$ with respect to $\bb$. For $\pol$, the number of total energy  evaluations depends on the dimension of the external field to construct the diagonal elements of the Hessian (Eq. \ref{eqn:polarizability_element}), which we also computed using AD. The Jacobian of $\pol$ with respect to $\bb$, a third order derivative, can be constructed from only three energy evaluations using AD \cite{ad_survey}, a drastic reduction from the 18 required for FD.

After implementation of the H\"uckel model using the \texttt{JAX} ecosystem \cite{jax2018github}, we could fully differentiate both observables, $\eHL$ and $\pol$. Importantly, using \texttt{JAX} allowed us to convert our existing Python-based H\"uckel code very easily by replacing calls to \texttt{NumPy} with almost equivalent calls to the \texttt{JAX.Numpy} package. For optimization, we used the Broyden–Fletcher–Goldfarb–Shanno ($\BFGS$) algorithm via the \texttt{JaxOpt} library \cite{jaxopt}. Instead of using a constrained optimization scheme to satisfy the site-normalization restriction for $\bb$, we used the softmax function,
\begin{equation}
    b^{i} = \frac{\exp(\underbar{b}^{i})}{\sum_{i}^M\exp(\underbar{b}^{i})},
    \label{eqn:softmax}
\end{equation}
where $\underbar{b}$ represents the unnormalized $b$ parameters.

\begin{figure}
    \centering
    \includegraphics[scale=0.45]{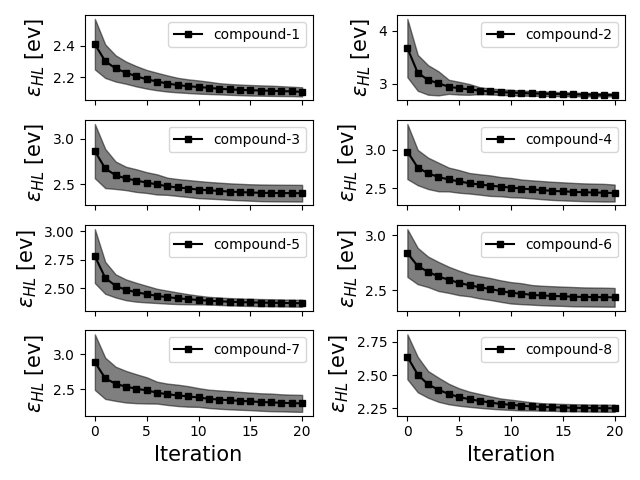}\label{fig:av_learning_curve_hl}
    \includegraphics[scale=0.45]{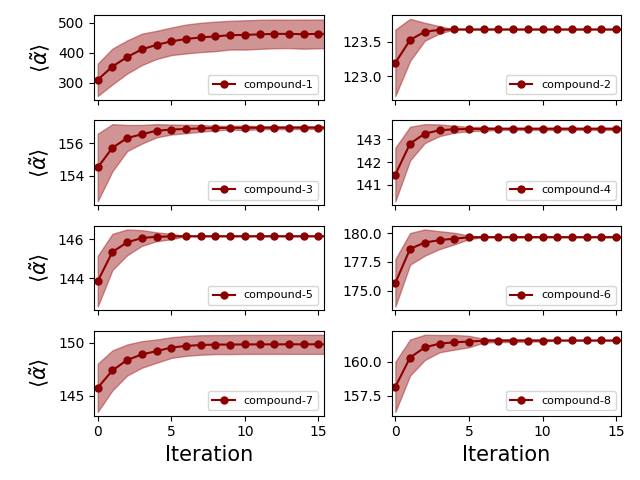}\label{fig:av_learning_curve_pol}
    \caption{Average learning curve for $\eHL$ (left panel) and $\pol$ (right panel) for 250 random initial molecules based on the eight different molecular frameworks. We use the $\BFGS$ algorithm to optimize both observables. More details about the random initialization are provided in the text.}
    \label{fig:av_learning_curve}
\end{figure}

\begin{figure}
    \centering
    \includegraphics[scale=0.35]{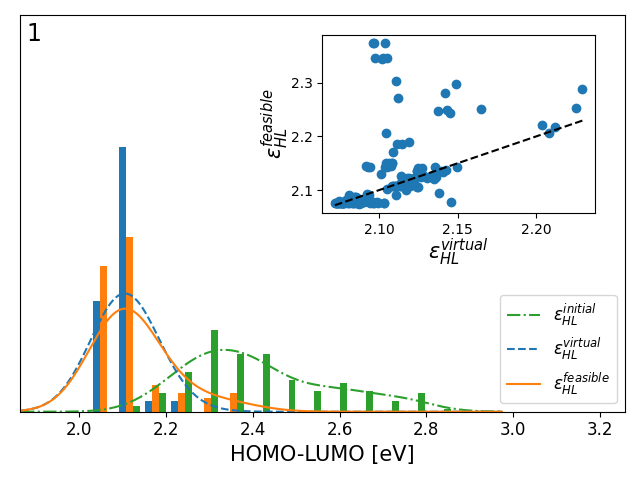}
    \includegraphics[scale=0.35]{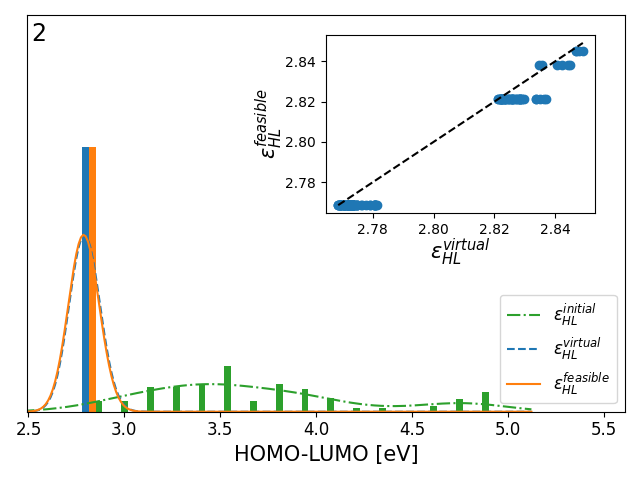}
    \includegraphics[scale=0.35]{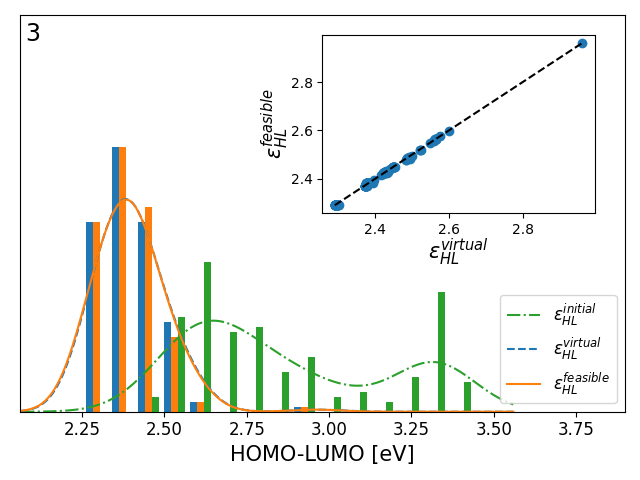}
    \includegraphics[scale=0.35]{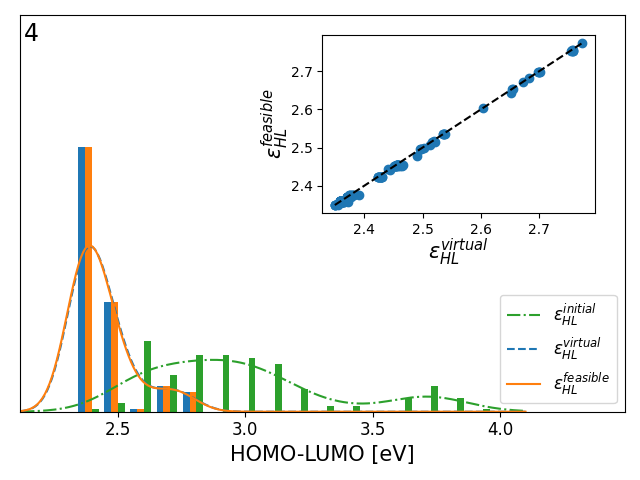}
    \includegraphics[scale=0.35]{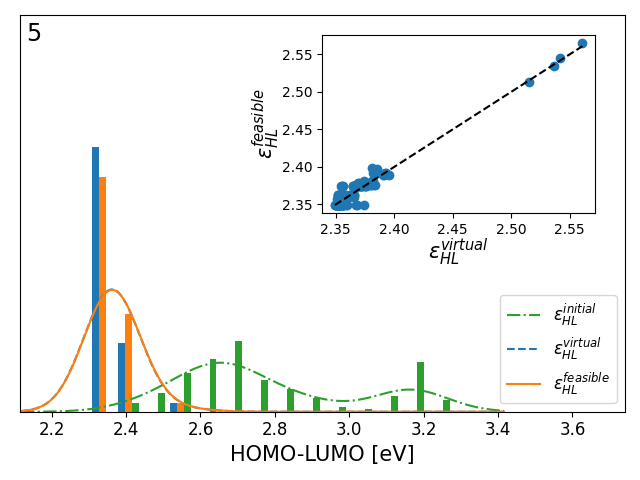}
    \includegraphics[scale=0.35]{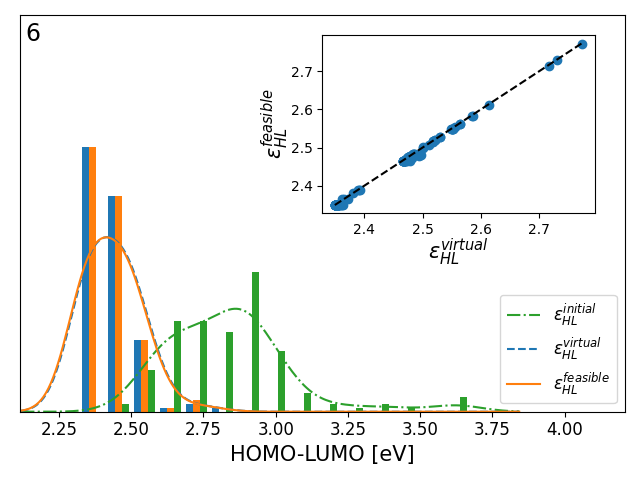}
    \includegraphics[scale=0.35]{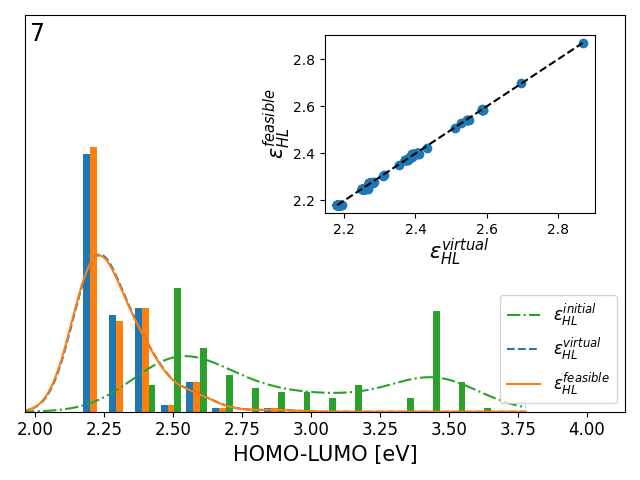}
    \includegraphics[scale=0.35]{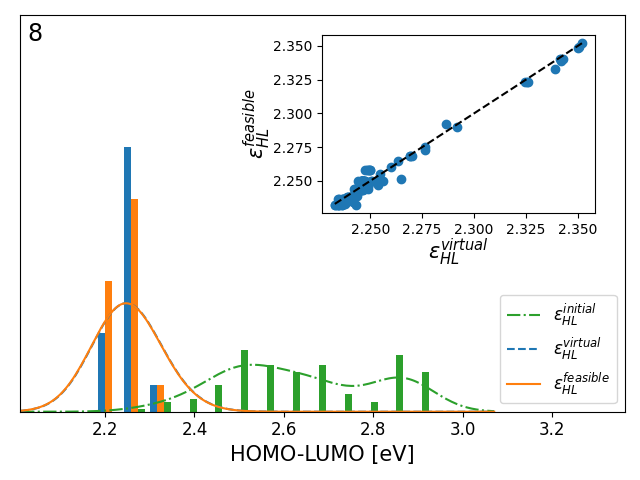}
    \caption{Histograms of the optimized HOMO-LUMO gap, Eq. \ref{eqn:homo_lumo}, for 250 random initial molecules ($\eHL^\mathrm{initial}$).
    The inset panels compare the similarity between $\eHL$ computed with the value of $\bb$ at the end of the optimization protocol ($\eHL^\mathrm{virtual}$), and the values of $\eHL$ selecting the most probable atoms given $\bb^{*}$ ($\eHL^\mathrm{feasible}$). Curves represent the derived histograms using kernel density estimation, (solid) $\eHL^\mathrm{feasible}$, (dashed) $\eHL^\mathrm{virtual}$, and (dotdash) $\eHL^\mathrm{initial}$.
    All molecules were optimized using the $\BFGS$ algorithm. 
    Molecular frameworks are displayed in Fig. \ref{fig:molecules}.
    }
    \label{fig:histogram_HL}
\end{figure}

\begin{figure}
    \centering
    \includegraphics[scale=0.35]{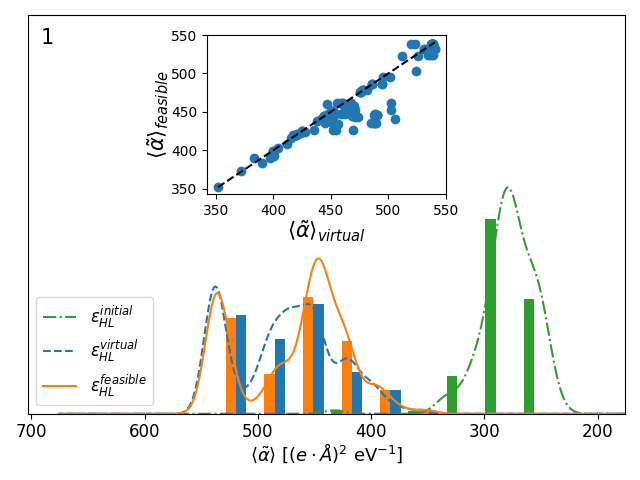}
    \includegraphics[scale=0.35]{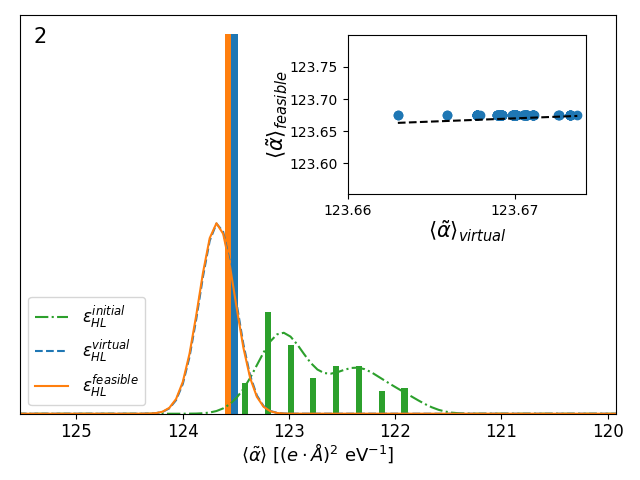}
    \includegraphics[scale=0.35]{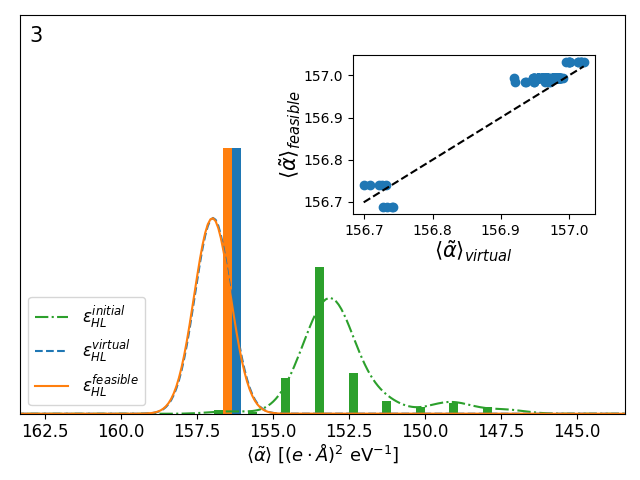}
    \includegraphics[scale=0.35]{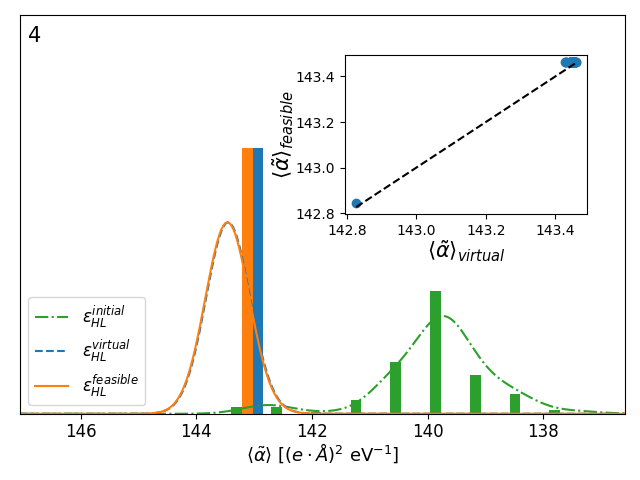}
    \includegraphics[scale=0.35]{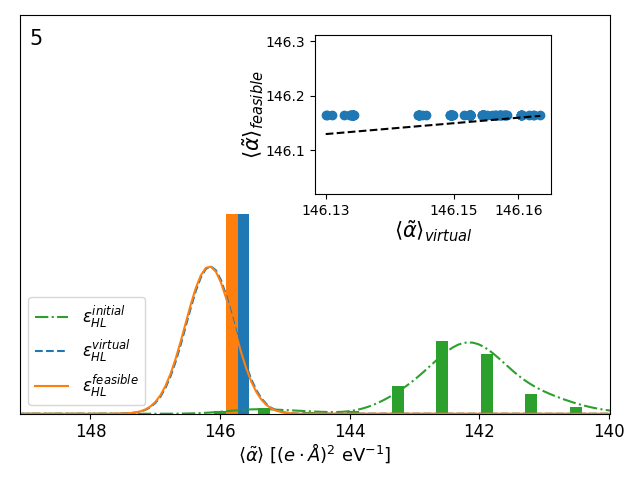}
    \includegraphics[scale=0.35]{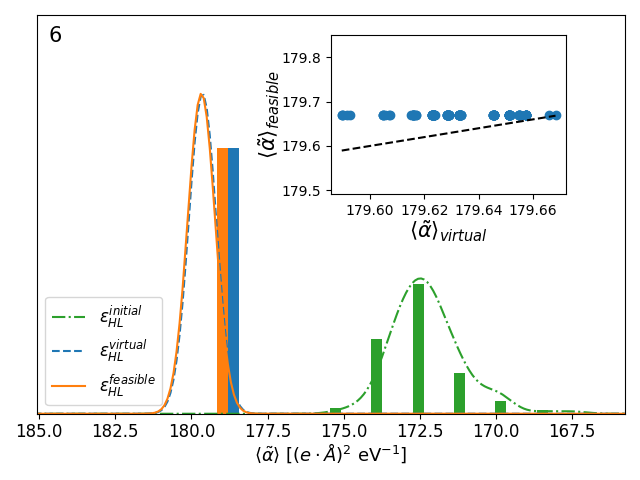}
    \includegraphics[scale=0.35]{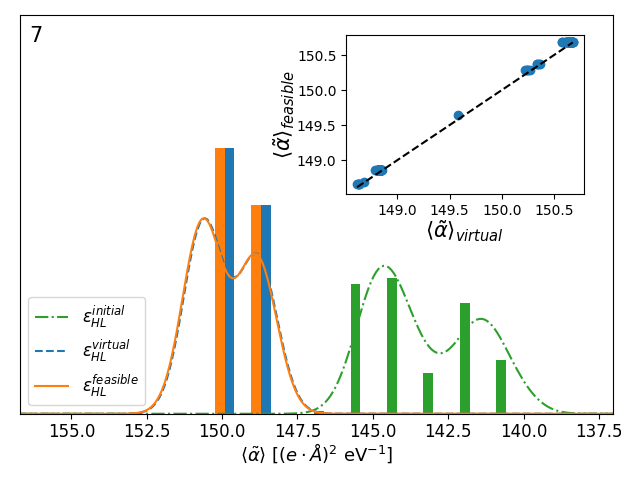}
    \includegraphics[scale=0.35]{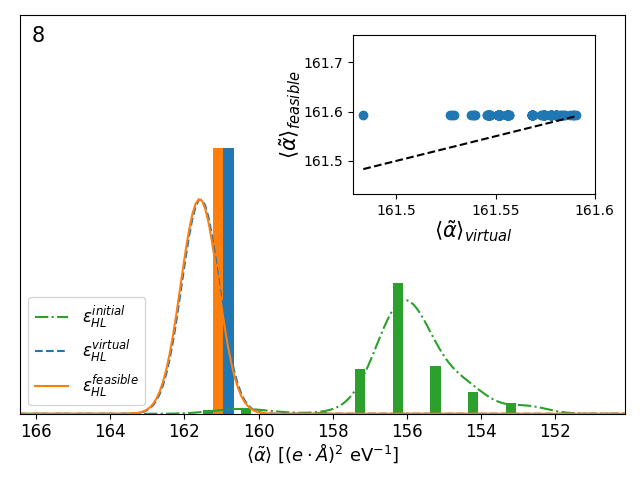}
    \caption{Histograms of the optimized $\pol$, Eq. \ref{eqn:polarizability}, for 250 random initial molecules ($\pol^\mathrm{initial}$).
    The inset panels compare the similarity between $\pol$ computed with the value of $\bb$ at the end of the optimization protocol ($\pol^\mathrm{virtual}$), and the values of $\pol$ selecting the most probable atoms given $\bb^{*}$ ($\pol^\mathrm{feasible}$). Curves represent the derived histograms using kernel density estimation, (solid) $\pol^\mathrm{feasible}$, (dashed) $\pol^\mathrm{virtual}$, and (dotdash) $\pol^\mathrm{initial}$.
    All molecules were optimized using the $\BFGS$ algorithm. 
    Molecular frameworks are displayed in Fig. \ref{fig:molecules}.
    }
    \label{fig:histogram_pol}
\end{figure}

Given the flexibility of the \texttt{JAX} ecosystem, we were able to test other gradient optimization algorithms such as \texttt{Adam} \cite{adamw} and canonical gradient-descent, but found the $\BFGS$ to be most efficient as it required, on average, fifteen or less total iterations to reach convergence (\textit{cf.} Fig. \ref{fig:av_learning_curve}). The \texttt{Adam} and gradient-descent algorithms, with an exponential learning rate decay, each needed more than thirty iterations to minimize $\eHL$ or $\pol$. Notably, we initialized the values for all $\underbar{b}^{i}$ parameters by sampling a uniform distribution, $\underbar{b}^{i}\sim {\cal U}(-1,1)$.
Instead of using literature Hückel parameters, we used our optimized parameters for each target observable where 5,000 training molecules were used. More details are described in Section \ref{sec:param_optimization}.
As seen in Figs. \ref{fig:histogram_HL} and \ref{fig:histogram_pol}, we found that our random initialization of $\bub$ allows us to sample a wide range of molecules with a broad range of values for both objectives, $\eHL^\mathrm{initial}$ and $\pol_\mathrm{initial}$.

\begin{figure}
    \centering
    \includegraphics[scale=0.5]{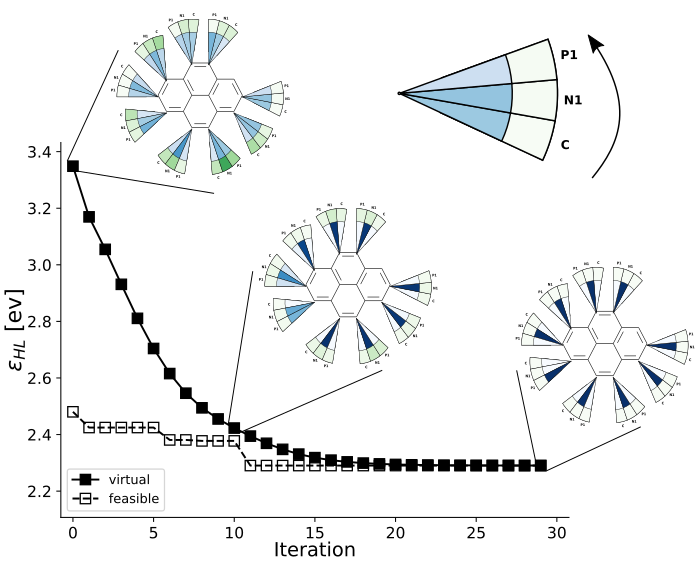}\\
    \includegraphics[scale=0.55]{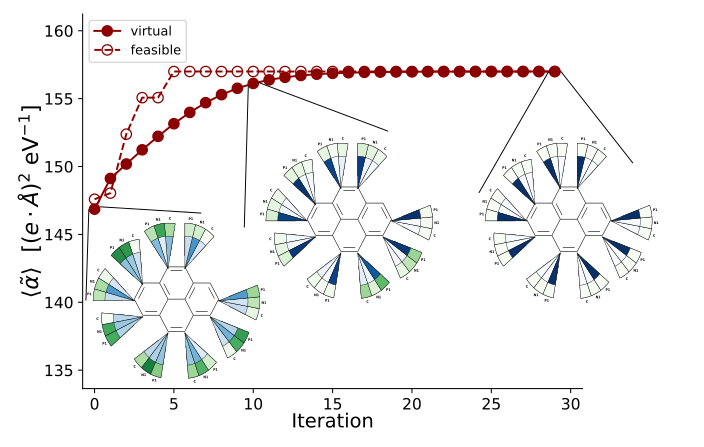}
    \caption{Change in the parameters $\bb$ for $\eHL$ (left panel) and $\pol$ (right panel) during the optimization of a single random initial virtual molecule based on framework \textbf{3}. For the initial, one intermediate and the final $\bb$, we plot the values of $\bb$ for molecular framework \textbf{3}. 
    For both panels, the filled markers represent the values of $\eHL$ and $\pol$ computed with $\bb$ at each iteration (virtual), and the empty markers represent the observable values computed only with the most probable atoms given $\bb$ at each iteration (feasible). 
    For each search site we only considered three different atom types, namely $C$, $N$ and $P$; $\bb_{\ell} = [b_{\ell}^{C},b_{\ell}^{N},b_{\ell}^{P}]$. The optimization of $\eHL$ and $\pol$ w.r.t. $\bb$ was carried out with \texttt{Adam} using a learning rate of 0.2. More details are provided in the main text.}
    \label{fig:learning_curve_HL_pol_molec_params}
\end{figure}

Because of the statistical description of the molecules by the $\bb$ parameters, the optimal parameters ($\bb^*$) found by optimization are not one-hot vectors that correspond to only one atom type per site in the molecule but rather a linear combination of multiple atom types. We define the observable value for this unphysical molecule as $y_\mathrm{virtual}$ and that for the real molecule as $y_\mathrm{feasible}$ (\textit{i.e.}, the most probable atom is picked for each site to define the real molecule). An example of this is displayed for framework \textbf{3} in Fig. \ref{fig:learning_curve_HL_pol_molec_params} where we show the change in $\bb$ throughout the optimization for both objectives ($\eHL$ and $\pol$). As we can observe, the change in $\eHL$ from the initial random molecule to the optimal one is close to 1~eV. The optimizations in Fig. \ref{fig:learning_curve_HL_pol_molec_params} were done using \texttt{Adam} only to properly illustrate the change in $\bb$ as the change between iterations is smoother and more readily discernable.
Importantly, this shows that the generative model can shift the distribution of properties towards the target property with little dependence on the random initialization of $\bb$ (Figs. \ref{fig:histogram_HL} and \ref{fig:histogram_pol}). We also observe that, for the majority of the optimized molecules, $y_{virtual}$ and $y_{feasible}$ are linearly correlated, indicating that the optimization converged essentially to feasible molecules. 
The property distributions of $y_\mathrm{virtual}$ and $y_\mathrm{feasible}$ are reasonably close, even in the few cases when the correlations are poor.

\begin{figure}
    \centering
    \includegraphics[scale=0.55]{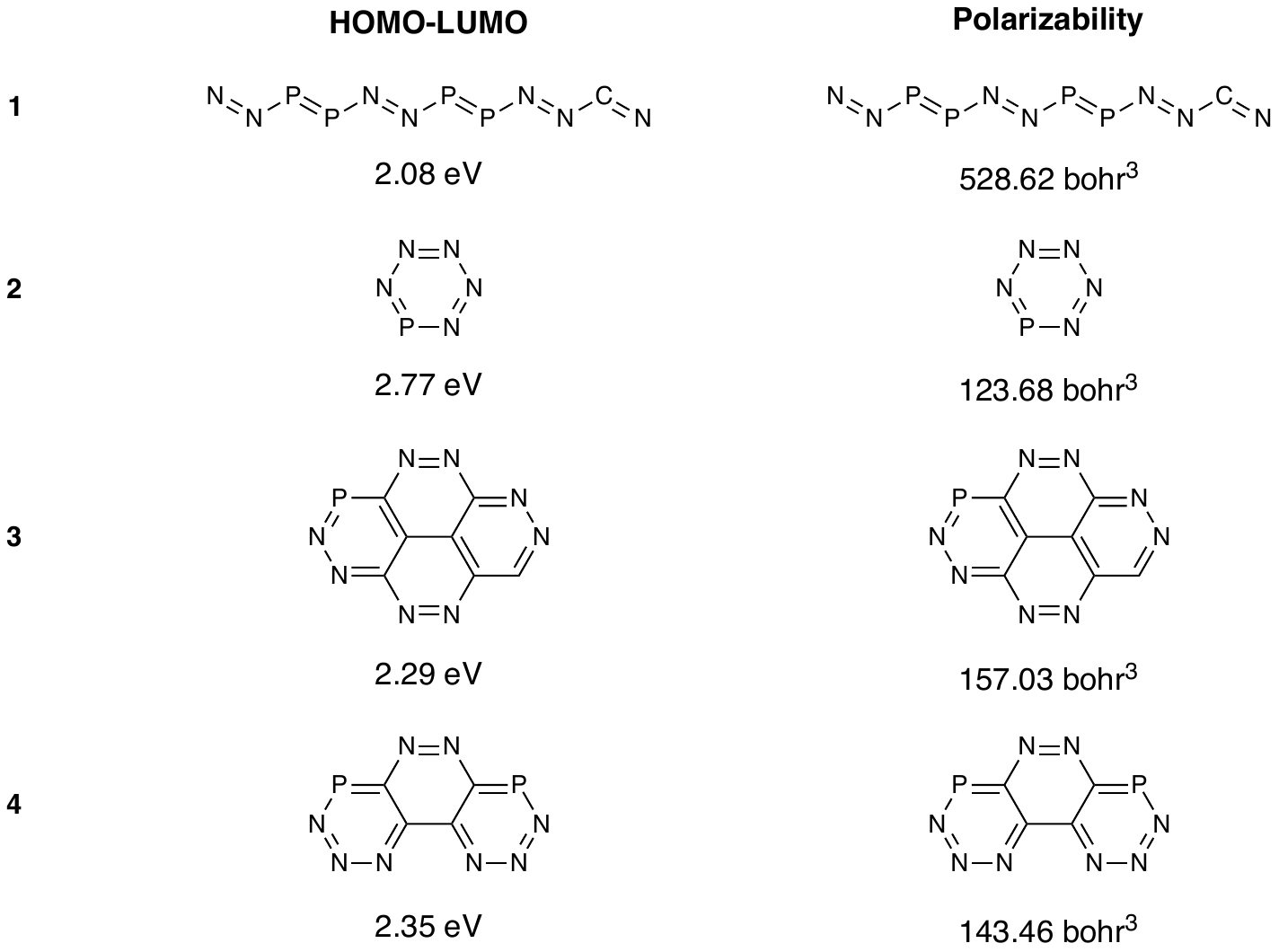} %
    \includegraphics[scale=0.50]{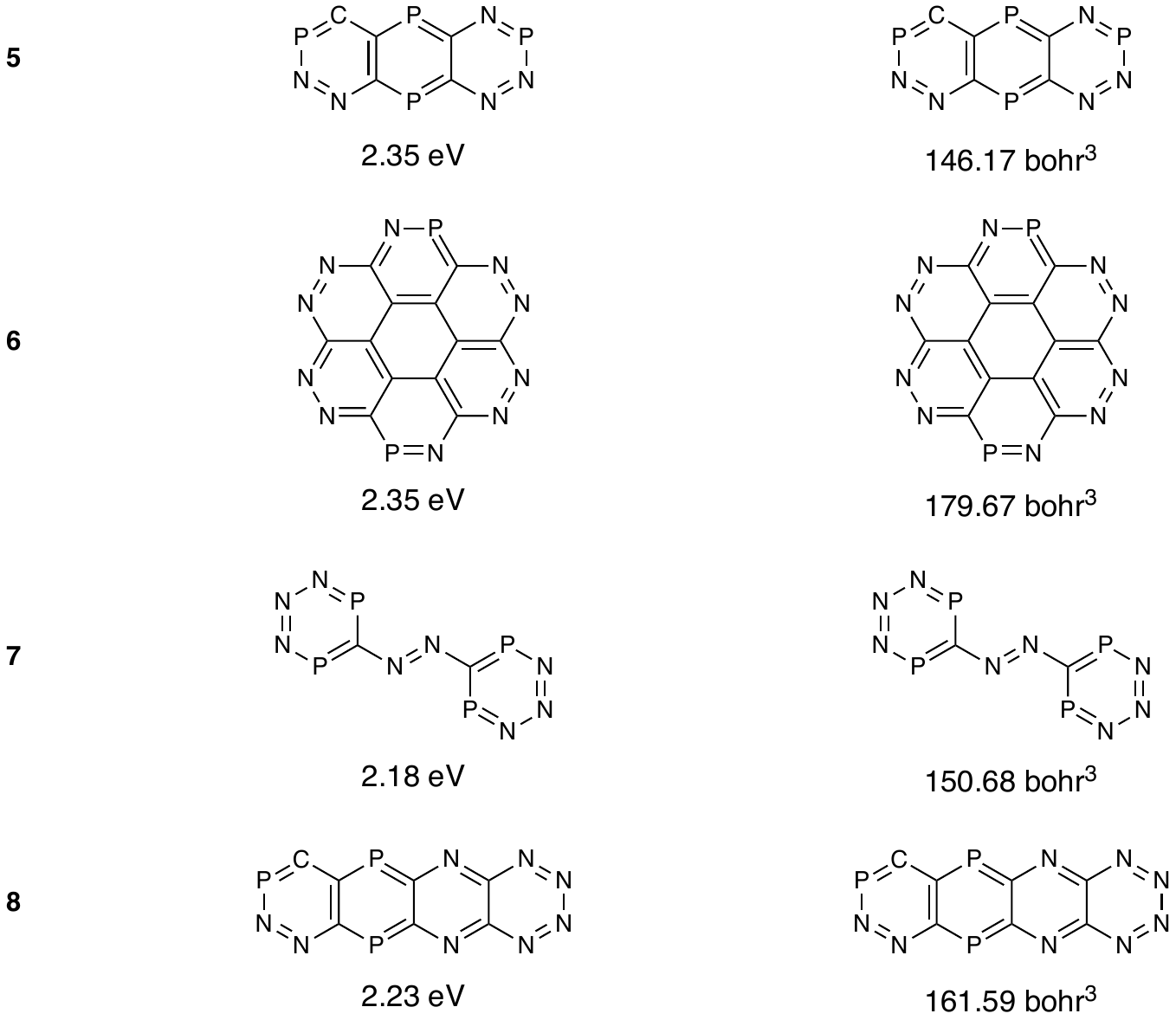}
    \caption{The molecular structures with the lowest HOMO-LUMO gap and maximum polarizability for each of the eighth molecules considered (Fig. \ref{fig:molecules}). }
    \label{fig:best_molecules}
\end{figure}

Fig. \ref{fig:best_molecules} displays the molecules with the lowest $\eHL$ and maximum $\pol$ from the ensemble of different optimizations. First, we notice that there is a higher amount of phosphorus atoms in the molecules when $\pol$ was the target property. This is not unexpected as molecular polarizabilities, while not simply a sum of the atomic polarizabilities, are strongly influenced by the atomic polarizabilities of the constituent atoms \cite{eisenlohr1911eine, steiger1921ein}. As phosphorus is a third-row element in the same group with nitrogen and atomic polarizability increases significantly when increasing the row number, its atomic polarizability, both in free atoms \cite{schwerdtfeger20192018} and in molecules \cite{Krawczuk2014polaber}, is significantly larger than both nitrogen and carbon. Therefore, incorporating a large number of phosphorus atoms is expected to be a viable strategy to maximize the molecular polarizability in all of the molecular frameworks considered. For the molecules with the lowest $\eHL$, we see extensive incorporation of both nitrogen and phosphorous atoms. This can be understood in terms of the effect of heteroatom substitution on $\eHL$ within the HMO framework \cite{heilbronner_1976}. For alternant compounds such as \textbf{1}--\textbf{8}, $\eHL$ is unaffected by changing $\alpha_{\ell}$. The main effect comes from changing $\beta_{\ell,k}$, and is expected to be largest for bonds that feature a bonding interaction in the HOMO and an anti-bonding interaction in the LUMO. A lowered $\beta_{\ell,k}$ leads to decreased bonding interactions in the HOMO and consequently, a higher HOMO energy. For the LUMO, decreasing the antibonding interactions by a lowered $\beta_{\ell,k}$ leads to a lowering of the energy. The net effect by raising the HOMO and lowering the LUMO is a decrease in the $\eHL$ gap. We can therefore expect optimization to favor atom pairs with a low $\beta_{\ell,k}$ for bonds that feature a bonding interaction in the HOMO and an antibonding interaction in the LUMO. Inspection of the molecular orbitals of the optimized frameworks (cf. Figure S1) indeed reveals that these bonds are dominantly between two N atoms, which feature the by far lowest $\beta_{\ell,k}$ at 0.159 (the next lowest is for P--P at 0.539). Control optimizations with the original parameter set by Van-Catledge \cite{van-catledgePariserParrPoplebasedSetHueckel1980} instead gives molecules with P--P at those bonds (cf. Figure S2), consistent with the fact that $\beta_{P,P}$ = 0.63 is the lowest for this parameter set.\\

For this proof of principle work, we picked two distinct molecular target properties. However, based on the framework employed, a significant number of alternative properties could also be predicted and, thus, used for inverse design via gradient-based optimization. Additionally, any combined objective that is derived from multiple target properties can equally be optimized for via the same types of algorithms out of the box. This is particularly interesting for properties where Hückel models are known to provide reasonable prediction accuracies such as HOMO-LUMO gaps. The use of gradient-based optimization algorithms enables fast convergence towards the closest local optimum solution reducing the number of evaluations and leading to significantly increased computation time. This is particularly important as one of the main bottlenecks in current approaches to inverse molecular design is the number of property evaluations needed to find an optimal structure \cite{gao2022sample}. Going beyond single-objective optimization, one possible extension of our presented approach is targeting multiple objectives via genuine gradient-based multi-objective optimization, for example, both $\eHL$ and $\pol$. The standard approach to perform multi-objective optimization is via
via property concatenation into a single function to use standard single-objective algorithms, where algorithms like Bayesian optimization are used \cite{chimera:AAG, bo_chem:AAG, Routescore:AAG, nanoparticle:AAG, mobopt:RAVH}. However, gradient-based multi-objective optimization algorithms \cite{MGDA} have been developed and they, together with automatic differentiation, could be employed for both parameter optimization and inverse molecular design in order to explore the corresponding Pareto fronts in a systematic manner. \\

From a conceptual point of view, representing chemical structure subspaces in a parameterized form can greatly facilitate inverse design \cite{Beratan_jcp_2008} as it allows the use of well-established approaches for parameter optimization to be used for the design of molecules. This is particularly effective when used in combination with AD due to its numerical stability and computational efficiency compared to alternative means to compute gradients. Consequently, this also makes the molecular size that can still be feasibly treated in such an approach larger and thus, essentially, expands the chemical subspace the generative model can explore. However, one of the main downsides of the approach implemented in this work is the reliance on fixed molecular frameworks, which is common for alchemical formulations \cite{rudorff2020alchemical} strongly limiting the structural space considered in the optimization. Simple extensions would be i) the combination of methods to change the molecular framework without relying on gradients with the method presented here to modify the atom identities within the respective framework, or ii) differentiable supermatrix structure where atom vacancies are allowed \cite{MKren_MLST}.
Ideally, future extensions should aim to find prudent ways allowing for framework modifications based on gradients as this potentially can lead to a dramatic reduction in the number of structure optimization steps and, thus, the number of property evaluations necessary.
The extended Hückel model is also compatible with the proposed methodology, even with ML learned parameters \cite{huckel:nn:jcp}, by considering a description of the overlap integrals between different atoms types, similar to Eq. \ref{eqn:tb_ham_invdesing}.

\subsection{Parameter optimization} \label{sec:param_optimization}

Another important task that might sometimes be underappreciated in computational chemistry is model parameter optimization. Here, we leverage the flexibility of AD and optimize all free parameters of the H\"uckel model in the same way as it is done for modern ML algorithms. Originally, the H\"uckel model is solely based on electronic interactions between nearest-neighbour atoms, which is typically also referred to as the tight-binding approximation (cf. $\beta_{\ell,k}$ parameters in Eq. \ref{eqn:tb_ham}). Beyond the standard Hückel model, one can introduce atomic distance-dependence of the corresponding interactions via  $\beta_{\ell,k} = \beta^{0}_{\ell,k}g(\mathbf{R}_{\ell,k})$. For example, based on previous work by Longuett-Higgins and Salem \cite{AlternationBondLengths1959}, $\beta_{\ell,k}$ has an exponential dependence on $\mathbf{R}_{\ell,k}$, 

\begin{equation}
    \beta_{\ell,k}^{exp} =  -\beta^{0}_{\ell,k} \exp^{-\frac{\Delta R_{\ell,k}}{y_{\ell,k}}} \label{eqn:beta_exp}.
\end{equation}

A second functional form, which is based on the work of Su, Schrieffer and Heege \cite{ssh:1979,ssh:1988} on conducting polymers, uses a linear distance-dependence of the interactions,

\begin{equation}
    \beta_{\ell,k}^{lr} = -\beta^{0}_{\ell,k}\left(1 - y^{-1}_{\ell,k}\Delta R_{\ell,k} \right ). \label{eqn:beta_linear}
\end{equation}

For both expressions (see Eqs. \ref{eqn:beta_exp}--\ref{eqn:beta_linear}), $\Delta R_{\ell,k}$ is the difference with respect to the reference bond length distance $R^{0}_{\ell,k}$, and $y_{\ell,k}$ is a length scale parameter. By including $R^{0}_{\ell,k}$ and $y_{\ell,k}$ in the set of parameters for the H\"uckel model, the complete set of parameters becomes $ \btheta =  [\alpha_\ell,\beta^{0}_{\ell,k},y_{\ell,k},R^{0}_{\ell,k}]$.

For this work, all initial $\alpha_\ell$ and $\beta^{0}_{\ell,k}$ parameters were taken from Van-Catledge \cite{van-catledgePariserParrPoplebasedSetHueckel1980}, and the initial $R^{0}_{\ell,k}$ parameters were approximated from tables of standard bond lengths \cite{r_xy_parameters}. The length scale parameters ($y_{\ell,k}$) were initially set to 0.3 $\text{\AA}$, which corresponds to the value that has been used for C--C in the literature \cite{AugHMO,AugHMOSSH}.

We used a subset of the GDB-13 data \cite{blum2009970} set that only consists of molecules with $\pi$-systems for fitting our model parameters (see Supplementary Material for details on how the dataset was generated). Note that some molecules in the dataset could have n-$\pi$* transitions as their lowest excited state. We used a pool of 60,000 molecules and randomly sampled 100, 1,000, and 5,000 molecules from this set, and used 1,000 additional molecules as validation set to monitor the optimization procedure.
To optimize $\btheta$, we used the mean squared error as a loss function,
\begin{equation}
    {\cal L}(\boldsymbol{\theta}) = \frac{1}{2} \sum_{i}^{N} \left ( \hat{\epsilon}_{HL}({\cal M}_{i}) - \eHL(\boldsymbol{\theta};{\cal M}_i) \right )^2,
    \label{eqn:loss}
\end{equation}
where ${\cal M}_i$ is a single molecule of the training set, and $\hat{\epsilon}_{HL}$ is the vertical excitation energy between the ground state and the first excited singlet state computed at the TDA-SCS-$\omega$PBEPP86/def2-SVP level of theory \cite{casanova2021time, weigend2005balanced}.
At the H\"uckel level of theory, this excitation energy simply corresponds to the HOMO-LUMO gap ($\eHL$) due to the disregard for electron correlation. To compare the prediction of $\eHL$ with the DFT reference values properly, we linearly transformed the results of the H\"uckel model using two additional parameters, $w_0$ and $w_1$ (Eq. \ref{eqn:linear_f_hm}), 

\begin{equation}
    \eHL(\boldsymbol{\theta};{\cal M}_i) = w_1\times \eHL(\alpha_\ell,\beta^{0}_{\ell,k},y_{\ell,k},R^{0}_{\ell,k};{\cal M}_i) + w_0, \label{eqn:linear_f_hm}
\end{equation}

where $\boldsymbol{\theta}$ jointly represents all parameters of the model, i.e., $\boldsymbol{\theta}=[\alpha_\ell,\beta^{0}_{\ell,k},y_{\ell,k},R^{0}_{\ell,k}, w_0,w_1]$.

For the optimization of all free parameters, we used the \texttt{AdamW} optimization algorithm \cite{adamw}, as implemented in the \texttt{Optax} library \cite{optax2020github}, with a learning rate of $0.02$, and a weight decay of $10^{-4}$.
Notably, we considered various training scenarios that included different values for the weight decay, and the regularization of different sets of the H\"uckel parameters. However, we found no impact on the accuracy of the model. The initial model parameters were gathered from Refs. \cite{van-catledgePariserParrPoplebasedSetHueckel1980,r_xy_parameters}. 

We optimized the parameters of three different H\"uckel models, i) the original one where $\beta_{\ell,k}$ is distance-independent ($\beta_{\ell,k} = \beta^{0}_{\ell,k} $), and both ii) the exponential (Eq. \ref{eqn:beta_exp}), and iii) the linear (Eq. \ref{eqn:beta_linear}) distance-dependence functional forms. We want to emphasize that any other analytic form for $\beta_{\ell,k}$ could be considered as well as AD makes any of these expressions fully differentiable. Following the convention in the literature, we scaled the parameters $\beta^{0}_{\ell,k}$ and $\alpha_{\ell}$ with respect to the carbon atom parameters according to $\alpha_{\ell} = \alpha_{\ell} - \alpha_{C}$, and $\beta^{0}_{\ell,k} = \beta^{0}_{\ell,k}/\beta^{0}_{C,C}$. Notably, at least for our results in this work, we found that including a regularization term in the loss function did not impact the accuracy of the model.  Finally, we found 20 epochs to be enough to minimize the loss function when the parameters are initialized with values from Refs. \cite{van-catledgePariserParrPoplebasedSetHueckel1980,r_xy_parameters}. \\

\begin{figure}
    \centering
    \includegraphics[scale=0.3]{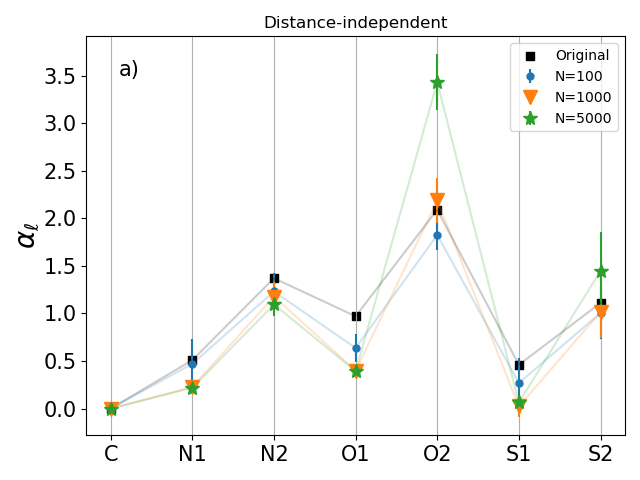}
    \includegraphics[scale=0.3]{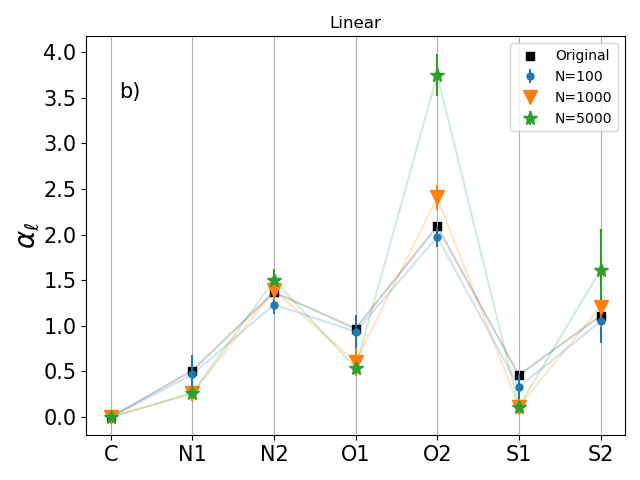}
    \includegraphics[scale=0.3]{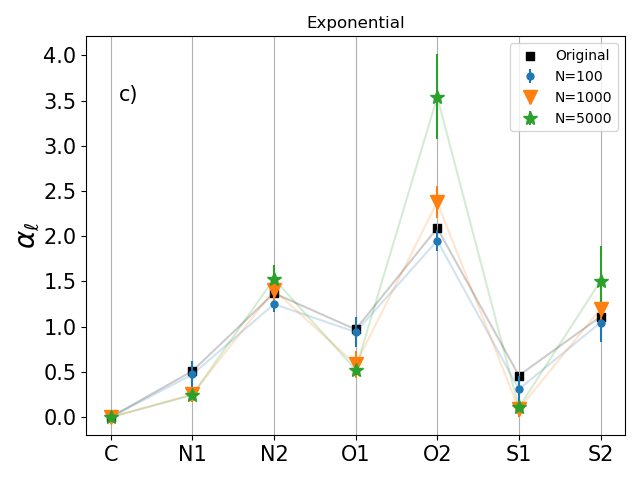}
    \caption{Optimized $\alpha_{\ell}$ parameters for different H\"uckel models, a) $\beta^{0}_{\ell,k}$, b) $\beta^{lr}_{\ell,k}$ (Eq. \ref{eqn:beta_linear}), and c) $\beta^{exp}_{\ell,k}$ (Eq. \ref{eqn:beta_exp}). All parameters reported were averaged over ten different training data sets, and different number of training molecules. Colored symbols and bars represent the mean and standard deviation of the optimized parameters averaged over ten different data sets. The reference parameters ($\blacksquare$-symbol) were taken from Refs. \cite{van-catledgePariserParrPoplebasedSetHueckel1980}, and used as the initial parameters for all optimizations. We refer the reader to the main text for the optimization details.}
    \label{fig:alpha_opt}
\end{figure}

\begin{figure}
    \centering
    \includegraphics[scale=0.3]{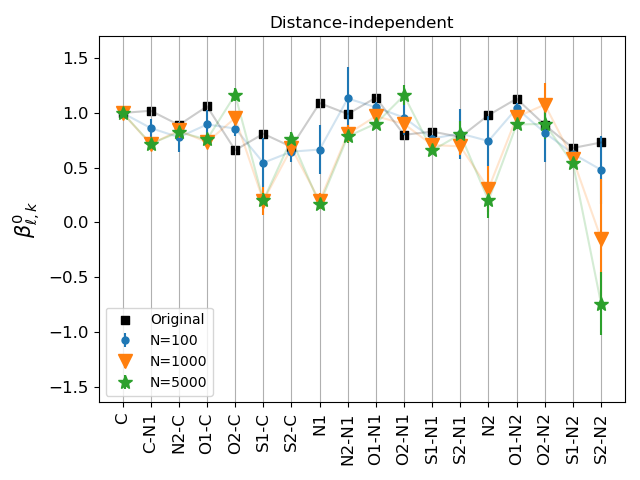}
    \includegraphics[scale=0.3]{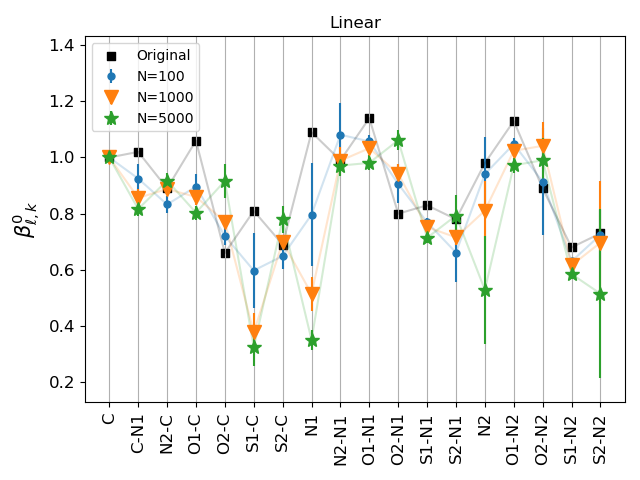}
    \includegraphics[scale=0.3]{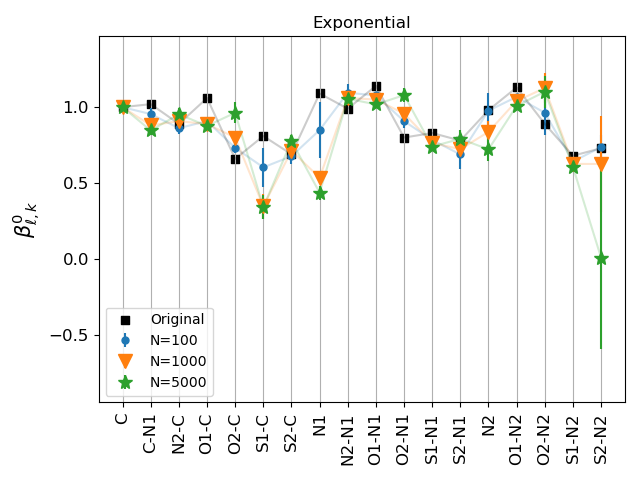}
    \caption{Optimized $\beta^{0}_{\ell,k}$ parameters for different H\"uckel models, a) $\beta^{0}_{\ell,k}$, b) $\beta^{lr}_{\ell,k}$ (Eq. \ref{eqn:beta_linear}), and c) $\beta^{exp}_{\ell,k}$ (Eq. \ref{eqn:beta_exp}). All parameters reported were averaged over ten different training data sets, and different number of training molecules. Colored symbols and bars represent the mean and standard deviation of the optimized parameters averaged over ten different data sets. The reference parameters ($\blacksquare$-symbol) were taken from Refs. \cite{van-catledgePariserParrPoplebasedSetHueckel1980}, and used as the initial parameters for all optimizations. We refer the reader to the main text for the optimization details.}
    \label{fig:beta_opt}
\end{figure}

\begin{figure}
    \centering
    \includegraphics[scale=0.3]{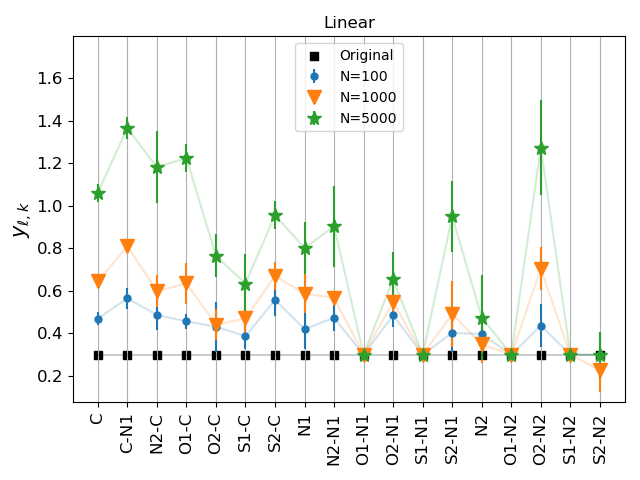}
    \includegraphics[scale=0.3]{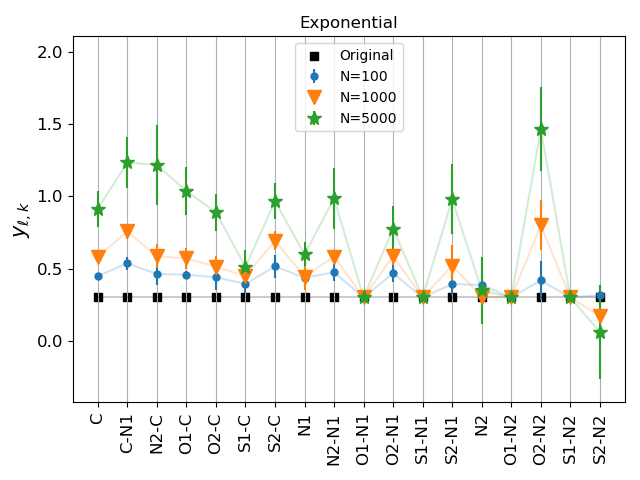}
    \caption{Optimized $y_{\ell,k}$ parameters for different H\"uckel models, a) $\beta^{lr}_{\ell,k}$ (Eq. \ref{eqn:beta_linear}), and b) $\beta^{exp}_{\ell,k}$ (Eq. \ref{eqn:beta_exp}). All parameters reported were averaged over ten different training data sets, and different number of training molecules. Colored symbols and bars represent the mean and standard deviation of the optimized parameters averaged over ten different data sets.The reference parameters ($\blacksquare$-symbol) were set to 0.3 $\text{\AA}$, and used as the initial parameters for all optimizations. We refer the reader to the main text for the optimization details.}
    \label{fig:param_y_opt}
\end{figure}

\begin{figure}
    \centering
    \includegraphics[scale=0.3]{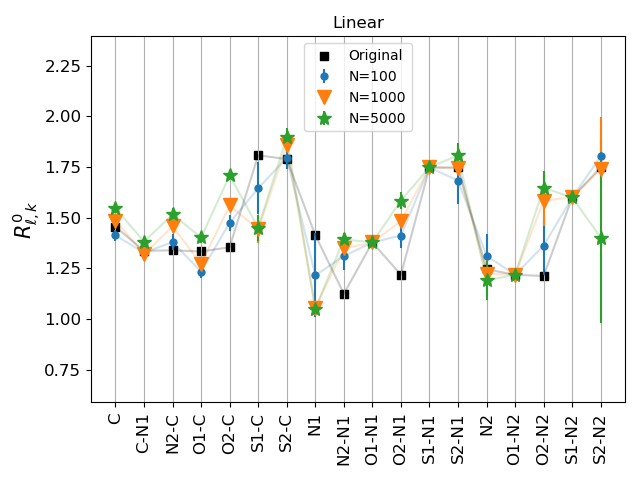}
    \includegraphics[scale=0.3]{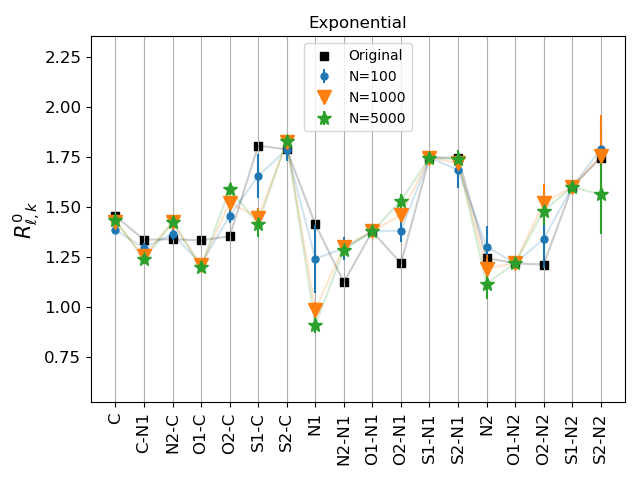}
    \caption{Optimized $R^{0}_{\ell,k}$ parameters for different H\"uckel models, a) $\beta^{lr}_{\ell,k}$ (Eq. \ref{eqn:beta_linear}), and b) $\beta^{exp}_{\ell,k}$ (Eq. \ref{eqn:beta_exp}). All parameters reported were averaged over ten different training data sets, and different number of training molecules. Colored symbols and bars represent the mean and standard deviation of the optimized parameters averaged over ten different data sets.
    The reference parameters ($\blacksquare$-symbol) were taken from Refs. \cite{r_xy_parameters}, and used as the initial parameters for all optimizations. We refer the reader to the main text for the optimization details.}
    \label{fig:param_R_opt}
\end{figure}

In Figures \ref{fig:alpha_opt}--\ref{fig:param_R_opt} we display the optimized values of the parameters for the three different H\"uckel models considered. From the optimized parameters, we observe that $\alpha_{O}$ (\textit{i.e.}, the 2p orbital energy parameter for oxygen), for all three models, is the one that differs the most from the literature \cite{van-catledgePariserParrPoplebasedSetHueckel1980,r_xy_parameters}. While there is no good reference data for $y_{\ell,k}$ to compare to, we observe, nevertheless, that the C--C parameter value changes considerably from the initial value of 0.3. For the $\beta^{0}_{\ell,k}$ parameters, only the values for N--C resemble the literature values. Furthermore, the optimal values of $R^{0}_{\ell,k}$ change the least from the values in Ref. \cite{r_xy_parameters}.

\begin{figure}
    \centering
    \includegraphics[scale=0.3]{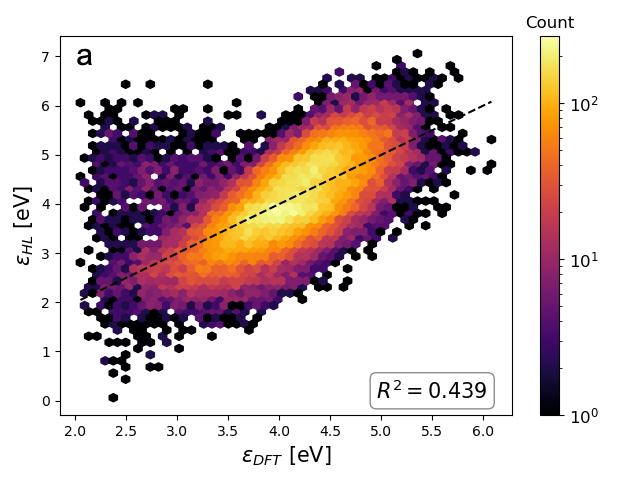}
    \includegraphics[scale=0.3]{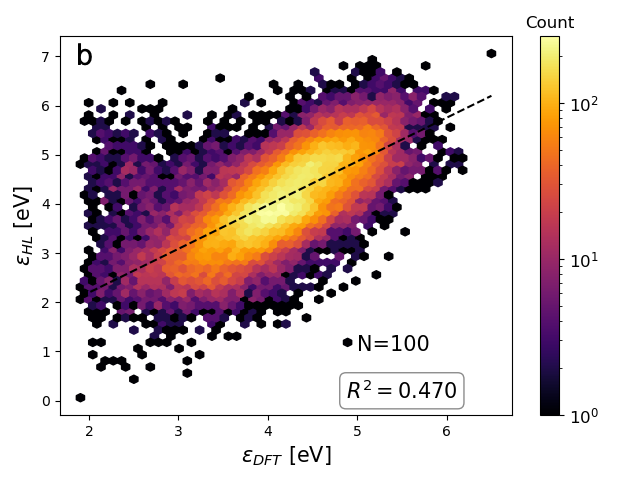}
    \includegraphics[scale=0.3]{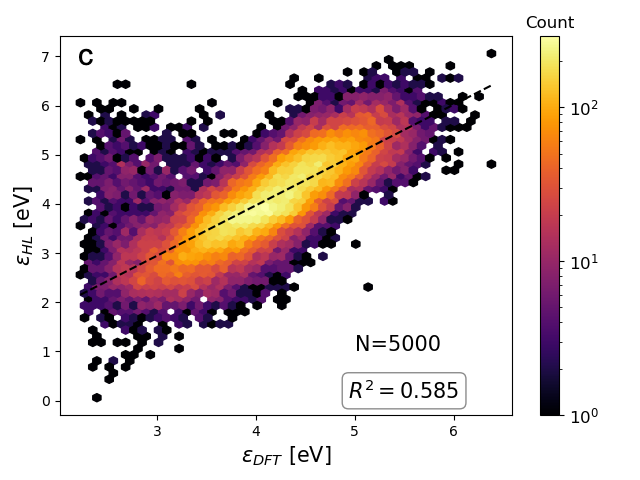}
    \caption{$\eHL$ predicted with the H\"uckel models and with DFT level for 40K molecules not considered during training. 
    The atom-atom interaction of the H\"uckel model is described by a distance-independent parameter, $\beta^{0}_{\ell,k}$. 
    Results computed with a H\"uckel with parameters taken from the literature (a), and parameters optimized with N=100 (b) and N=5,000 (c) data points. We refer the reader to the main text for the optimization details.}
    \label{fig:pred_beta_c}
\end{figure}

\begin{figure}
    \centering
    \includegraphics[scale=0.3]{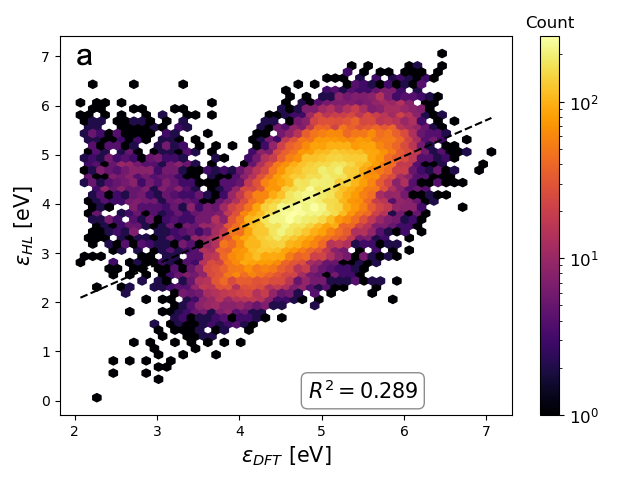}
    \includegraphics[scale=0.3]{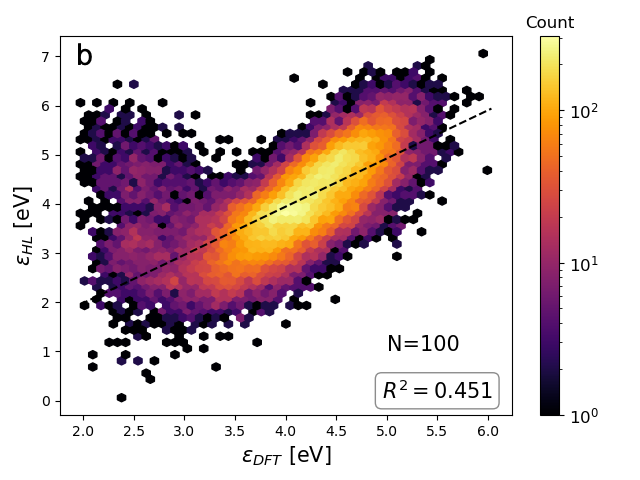}
    \includegraphics[scale=0.3]{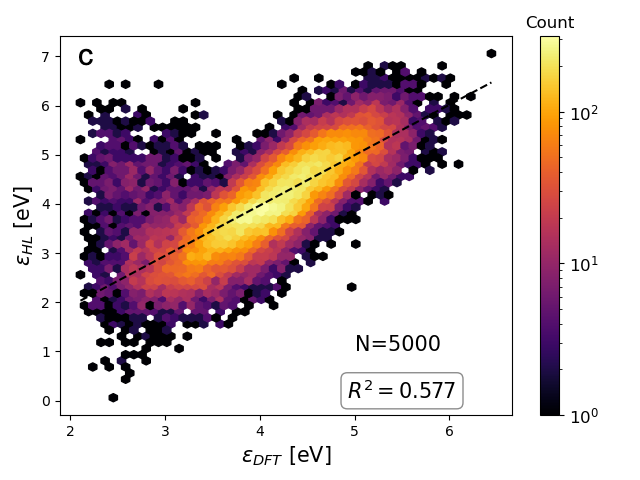}
    \caption{$\eHL$ predicted with the H\"uckel models and with DFT level for 40K molecules not considered during training. 
    The atom-atom interaction of the H\"uckel model is described by a distance-dependent parameter, $\beta^{lr}_{\ell,k}$ (Eq. \ref{eqn:beta_linear}). 
    Results computed with a H\"uckel with parameters taken from the literature (a), and parameters optimized with N=100 (b) and N=5,000 (c) data points. We refer the reader to the main text for the optimization details.}
    \label{fig:pred_beta_linear}
\end{figure}

\begin{figure}
    \centering
    \includegraphics[scale=0.3]{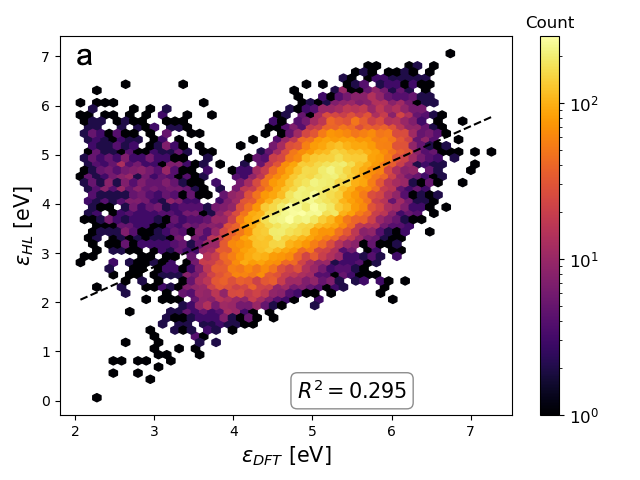}
    \includegraphics[scale=0.3]{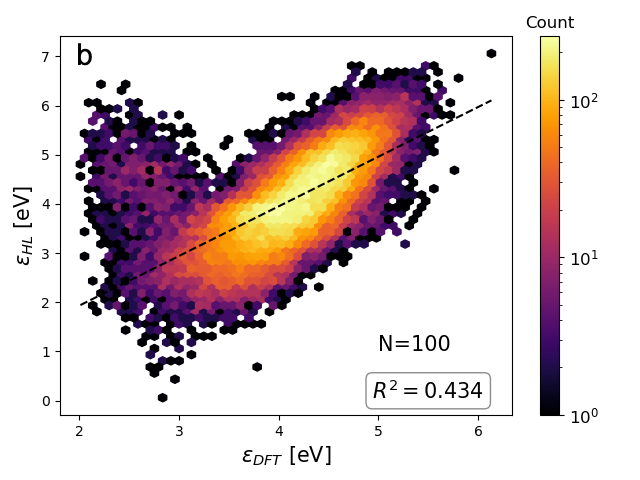}
    \includegraphics[scale=0.3]{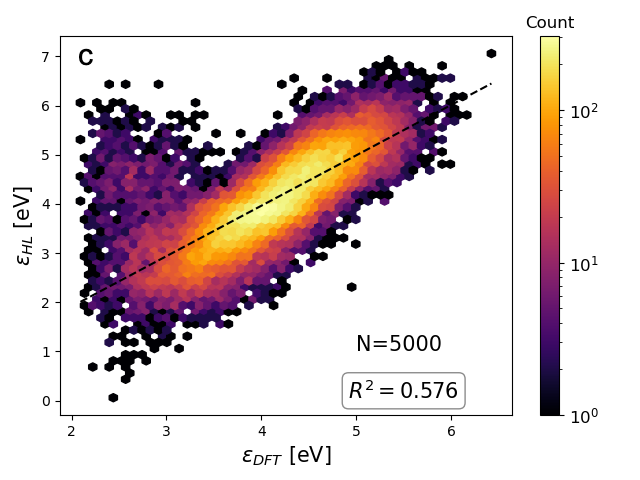}
    \caption{$\eHL$ predicted with the H\"uckel models and with DFT level for 40K molecules not considered during training. 
    The atom-atom interaction of the H\"uckel model is described by a distance-dependent parameter, $\beta^{exp}_{\ell,k}$ (Eq. \ref{eqn:beta_exp}). 
    Results computed with a H\"uckel model with parameters taken from the literature (a), and parameters optimized with N=100 (b) and N=5,000 (c) data points. We refer the reader to the main text for the optimization details.}
    \label{fig:pred_beta_exp}
\end{figure}

Using these optimized parameters, we predicted $\eHL$ for 40,000 additional test set molecules and compare the results with DFT reference data. The direct comparison is depicted in Figures \ref{fig:pred_beta_c}--\ref{fig:pred_beta_exp}. By optimizing all parameters there is a significant improvement in the prediction of $\eHL$ using our semi-empirical model. Notably, we also found that considering a larger training data set does not impact the accuracy of the H\"uckel model which suggests that either the corresponding molecules do not provide any additional information with respect to the relevant interaction parameters or that the model already is close to its best expected performance and cannot be improved further. Another important observation in that regard is that the analytical form of $g(\mathbf{R}_{\ell,k})$ in $\beta_{\ell,k}$ does not impact the accuracy of the model when optimized parameters are used.

\begin{figure}
    \centering
    \includegraphics[scale=0.3]{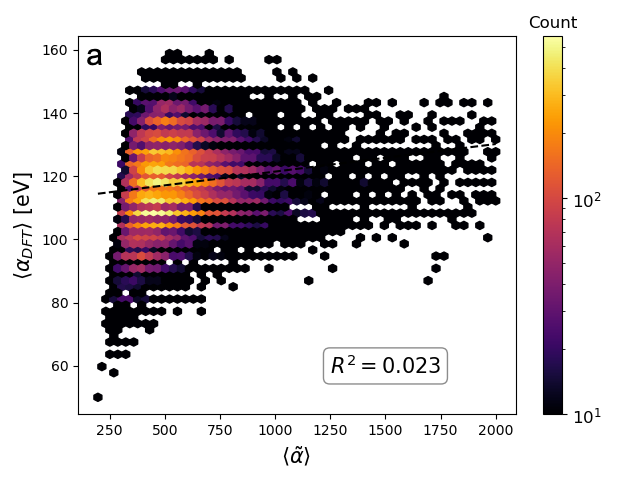}
    \includegraphics[scale=0.3]{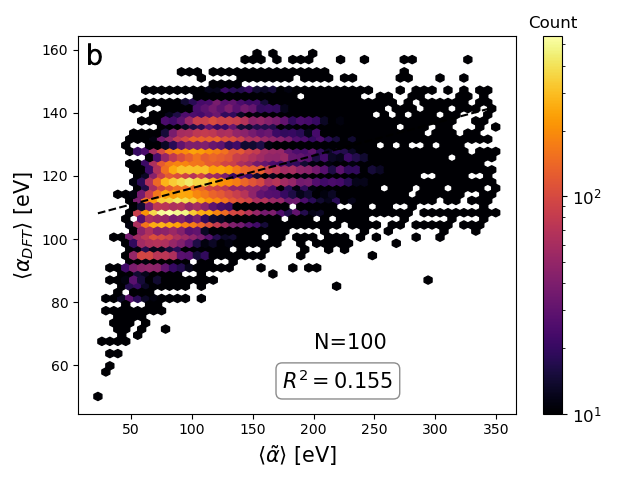}
    \caption{$\pol$ predicted with the H\"uckel models and with DFT level for 40K molecules not considered during training. 
    The atom-atom interaction of the H\"uckel model is described by a distance-independent parameter, $\beta^{0}_{\ell,k}$. 
    Results computed with a H\"uckel model with parameters taken from the literature (a), and (b) parameters optimized with N=100 data  points. We refer the reader to the main text for the optimization details.}
    \label{fig:pred_pol}
\end{figure}

Next, we also optimized the parameters with respect to the polarizability for the distance-independent H\"uckel model.  Here, the gradients needed for training are of third-order, e.g., $ \frac{\partial }{\partial \alpha_\ell} \frac{\partial^2 \tilde{\alpha}_{kk}}{\partial F^2_{k} }$ or $ \frac{\partial }{\partial \beta_{k,\ell}} \frac{\partial^2 \tilde{\alpha}_{kk}}{\partial F^2_{k} }$, and can be computed more efficiently via AD, illustrating the potential of this approach.
The molecular polarizabilities that were used as reference data were computed using \texttt{dftd4} (version 3.4.0) \cite{grimme:d3,grimme:london,grimme:d4} via the default methodology summing atomic polarizabilities.
Even though we observe a higher prediction accuracy when the optimized parameters are used compared to the model before parameter refinement, as depicted in Fig. \ref{fig:pred_pol}, the accuracy of the model still remains relatively low and does not improve anymore when more training data is used. We suspect that this prediction task is particularly challenging for the Hu\"ckel model as the simulated polarizability only corresponds to the contribution from $\pi$-electrons, while that of the reference data accounts for all the electrons in the molecules. Even though we expect a significant portion of the molecular polarizabilities to stem from the $\pi$-electrons, the contributions of the $\sigma$-electrons cannot be neglected and can dominate this property. Nevertheless, this proof-of-concept application example demonstrates the operational ease of conducting parameter refinement of a given physics-based prediction model based on reference data, even when derivative properties are targeted. \\


\break
\section{Conclusions}
In this work, we demonstrate the power of automatic differentiation to enable the efficient use of physics-inspired models for gradient-based optimization problems in the realm of molecular chemistry via semi-empirical Hückel models. In particular, we showcase inverse molecular design via an alchemical problem formulation using fixed molecular frameworks. This allows us to perform structure optimization requiring only a very small number of intermediate structures to find local minima with respect to the properties of interest utilizing gradients with respect to atom identities at specific sites. While our approach is currently limited to a fixed molecular framework, performing optimizations over the molecular composition space alone is far from trivial. Compared to various alternative approaches, our implementation shows a remarkably high molecular sampling efficiency due to efficient utilization of gradient information in combination with powerful gradient-based optimization algorithms. Additionally, we showcase the ease of generating calibrated physics-based property prediction models using high quality reference training data of relatively modest size, again allowing for quick convergence of model parameters. This is particularly important as most physical models that rely on empirical parameters such as semi-empirical quantum chemistry models and density functional approximations are still largely optimized by hand, making the corresponding procedures tedious. Thus, we believe that our work will serve as an inspiration for the field of computational chemistry in order to adopt the readily available AD capabilities of mature ML programming frameworks allowing to accelerate the construction of ever more accurate physics-based property simulation models.

\break
\section*{Acknowledgments}
 R.P. acknowledges funding through a Postdoc.Mobility fellowship by the Swiss National Science Foundation (SNSF, Project No. 191127).
 A.A.-G. thanks Anders G. Fr\o{}seth for his generous support. A.A.-G. acknowledges the generous support of Natural Resources Canada and the Canada 150 Research Chairs program. We also thank the SciNet HPC Consortium for support regarding the use
of the Niagara supercomputer. SciNet is funded by the Canada Foundation for Innovation, the Government
of Ontario, Ontario Research Fund - Research Excellence, and the University of Toronto. 

\section*{Data Availability}
The data that support the findings of this study are available within the article and its supplementary material and in GitHub 
(inverse molecular design) \url{https://github.com/RodrigoAVargasHdz/huxel_molecule_desing} and (parameter optimization) \url{https://github.com/RodrigoAVargasHdz/huxel}. 
\break
\section*{References}
\bibliography{references} 

\end{document}


\renewcommand{\thepage}{S\arabic{page}}
\renewcommand{\thesection}{S\arabic{section}}
\renewcommand{\thetable}{S\arabic{table}}
\renewcommand{\thefigure}{S\arabic{figure}}

\title[]{Supplemental material for "Inverse molecular design and parameter optimization with H\"uckel theory using automatic differentiation"}

\author{Rodrigo A. Vargas--Hern\'andez}
\address{Chemical Physics Theory Group, Department of Chemistry, University of Toronto, Toronto, Ontario, M5S 3H6, Canada.}
\address{Vector Institute for Artificial Intelligence, 661 University Ave. Suite 710, Toronto, Ontario M5G 1M1, Canada.}
\address{current affiliation: Department of Chemistry and Chemical Biology, McMaster University, 1280 Main Street West, Hamilton, Ontario, L8S 4M1 Canada.}
\ead{vargashr@mcmaster.ca}

\author{Kjell Jorner}
\address{Chemical Physics Theory Group, Department of Chemistry, University of Toronto, Toronto, Ontario, M5S 3H6, Canada.}
\address{Department of Computer Science, University of Toronto, 40 St. George St, Ontario M5S 2E4, Canada.}
\address{Department of Chemistry and Chemical Engineering, Chalmers University of Technology, Kemigården 4, SE-41258, Gothenburg, Sweden.}

\author{Robert Pollice}
\address{Chemical Physics Theory Group, Department of Chemistry, University of Toronto, Toronto, Ontario, M5S 3H6, Canada.}
\address{Department of Computer Science, University of Toronto, 40 St. George St, Ontario M5S 2E4, Canada.}
\address{current affiliation: Stratingh Institute for Chemistry, University of Groningen, Nijenborgh 4, Groningen, 9747 AG, The Netherlands.}

\author{Al\'an Aspuru--Guzik}
\address{Department of Chemistry, Chemical Physics Theory Group, 80 St. George St., University of Toronto, Ontario M5S 3H6, Canada.}
\address{Department of Computer Science, University of Toronto, 40 St. George St, Ontario M5S 2E4, Canada.}
\address{Department of Chemical Engineering \& Applied Chemistry, 200 College St., University of Toronto, Ontario M5S 3E5, Canada.}
\address{Department of Materials Science \& Engineering, 184 College St., University of Toronto, Ontario M5S 3E4, Canada.}
\address{Vector Institute for Artificial Intelligence, 661 University Ave. Suite 710, Toronto, Ontario M5G 1M1, Canada.}
\address{Lebovic Fellow, Canadian Institute for Advanced Research (CIFAR), 661 University Ave., Toronto, Ontario M5G 1M1, Canada.}

\vspace{10pt}



\section{GDB-13 Subset}
We developed a comprehensive set of filters to define a subset of GDB-13 \cite{blum2009970}, originally comprising more than 970 million organic molecules, with over 400,000 structures possessing cycles and a high degree of conjugation. These filters ensure that the molecules encompass cyclic $\pi$-systems to increase representation of the relevant structural space that is appropriate for simulation via the Hückel method. They were implemented via RDKit \cite{landrum2006rdkit} and are summarized in Table \ref{table_gdb13_subset}. For the forbidden substructures, we utilized the following SMARTS patterns:

\begin{lstlisting}[basicstyle=\ttfamily\small,breaklines]
'[Cl,Br,I]'
'*=*=*'
'*#*'
'[O,o,S,s]~[O,o,S,s]'
'[N,n,O,o,S,s]~[N,n,O,o,S,s]~[N,n,O,o,S,s]'
'[C,c]~N=,:[O,o,S,s;!R]'
'[N,n,O,o,S,s]~[N,n,O,o,S,s]~[C,c]=,:[O,o,S,s,N,n;!R]'
'*=[NH]'
'*=N-[*;!R]'
'*~[N,n,O,o,S,s]-[N,n,O,o,S,s;!R]'
'*-[CH1]-*'
'*-[CH2]-*'
'*-[CH3]'
\end{lstlisting}

\setlength{\tabcolsep}{16pt}
\begin{table*}[htbp!]
\caption{List of filters employed to create the $\pi$-systems subset of GDB-13. In each line, $x$ denotes the value of the corresponding feature.}
\label{tab:gdb13_subset_filters}
\begin{center}
\resizebox{1.\textwidth}{!}{
\begin{tabular}{cclc}
\hline
\textbf{Number} & \textbf{Feature} & \textbf{Definition} & \textbf{Value} \\ 
\hline
1 & Charge & Charge of the molecule. & $x = 0$ \\ 
2 & Radicals & Number of radical electrons. & $x = 0$ \\ 
3 & Bridgehead atoms & Number of bridgehead atoms. & $x = 0$ \\ 
4 & Spiro atoms & Number of spiro atoms. & $x = 0$ \\ 
5 & Aromaticity degree & Percentage of aromatic non-hydrogen atoms. & $x \geq 0.5$ \\ 
6 & Conjugation degree & Percentage of conjugated bonds between non-hydrogen atoms. & $x \geq 0.7$ \\ 
7 & Maximum ring size & Size of the largest ring. & $4 \leq x \leq 8$ \\ 
8 & Minimum ring size & Size of the smallest ring. & $4 \leq x \leq 8$ \\ 
9 & Substructures & Presence of any forbidden substructures (see text). & FALSE \\
\hline
\label{table_gdb13_subset}
\end{tabular}}
\end{center}
\end{table*}

On the structures obtained after filtering, we performed property simulations. To simulate the excited state properties of all the molecules, first, ground state conformational ensembles were generated using \texttt{CREST} \cite{pracht2020conformer} (version 2.11) with the iMTD-GC \cite{grimme2019exploration, pracht2020automated} workflow (default option) via the GFN2-xTB//GFN-FF \cite{bannwarthGFN2xTBAccurateBroadly2019, spicherRobustAtomisticModeling2020} composite method. The lowest energy conformers were optimized using \texttt{ORCA} \cite{neese2022software} (version 5.0.1) at the r2SCAN-3c \cite{grimme2021r2scan} level of theory. The corresponding geometries were used for subsequent ground- and excited state single-point calculations to assess the properties of vertical excitations. Single points at the TDA-SCS-$\omega$PBEPP86/def2-SVP \cite{casanova2021time, weigend2005balanced} level of theory were performed using \texttt{ORCA} (version 5.0.1) accounting for 8 roots in the singlet excited state manifold. The resolution of identity (RI) approximation was used via the RIJCOSX approach \cite{neese2009efficient, izsak2011an} for the exchange-correlation potential as implemented in \texttt{ORCA} using the def2/J \cite{weigend2006accurate} auxiliary basis set, and for the RI-MP2 \cite{feyereisen1993use, weigend1997ri, distasio2007an} component of the simulation via the def2-SVP/C auxiliary basis set.

\begin{figure}
    \centering
    \begin{subfigure}[a]{0.49\textwidth}
        \centering
        \textsf{HOMO}
    \end{subfigure}
    \begin{subfigure}[a]{0.49\textwidth}
        \centering
        \textsf{LUMO}
    \end{subfigure}
    \vspace{0.5cm}        
    
    {\small\textsf{\textbf{1}}}
    \begin{subfigure}[a]{0.48\textwidth}
        \centering
        \includegraphics[height=0.9cm]{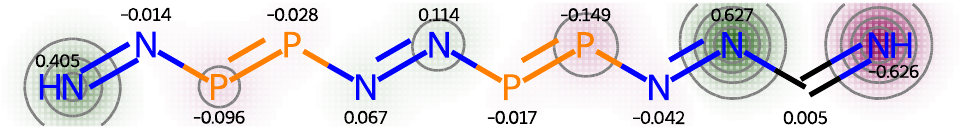}
    \end{subfigure}
    \begin{subfigure}[a]{0.48\textwidth}
        \centering
        \includegraphics[height=0.9cm]{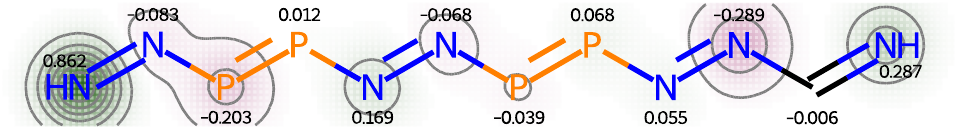}
    \end{subfigure}
    {\small\textsf{\textbf{2}}}
    \begin{subfigure}[a]{0.48\textwidth}
        \centering
        \includegraphics[height=2cm]{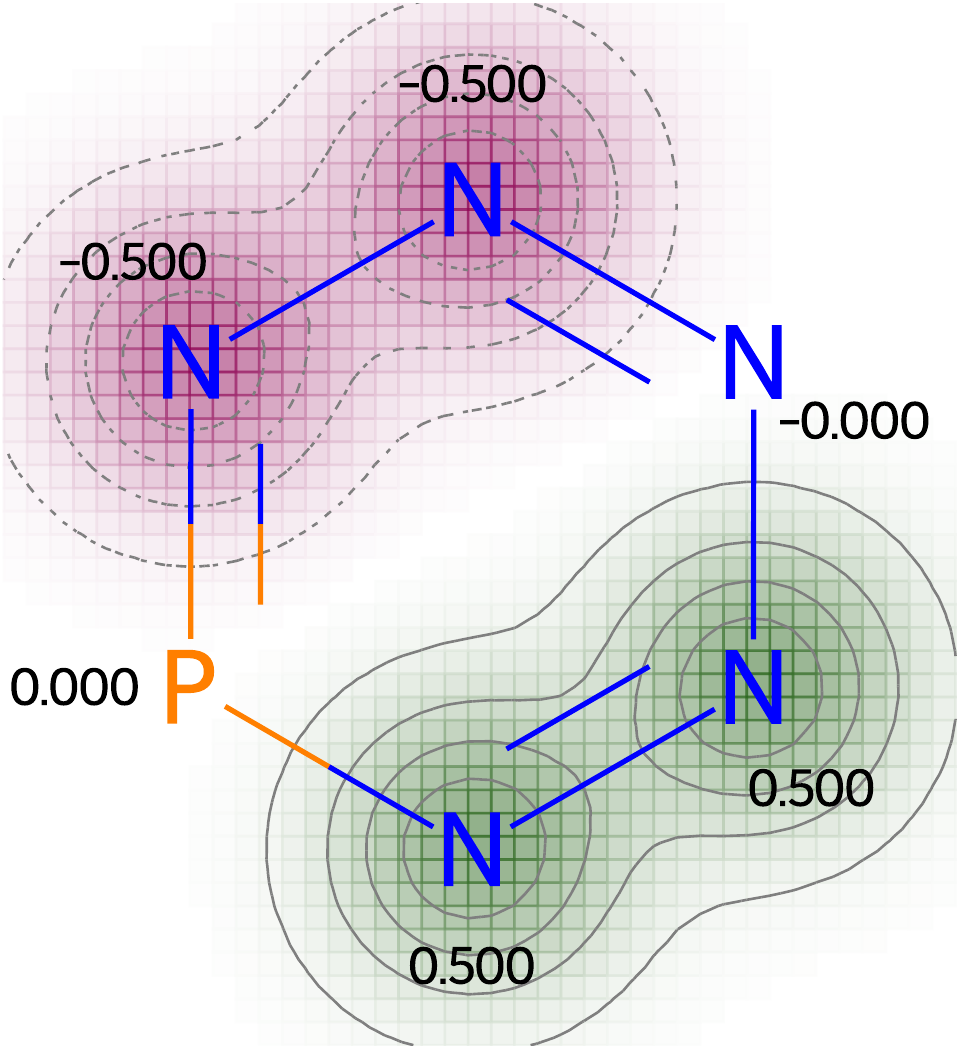}
    \end{subfigure}
    \begin{subfigure}[a]{0.48\textwidth}
        \centering
        \includegraphics[height=2cm]{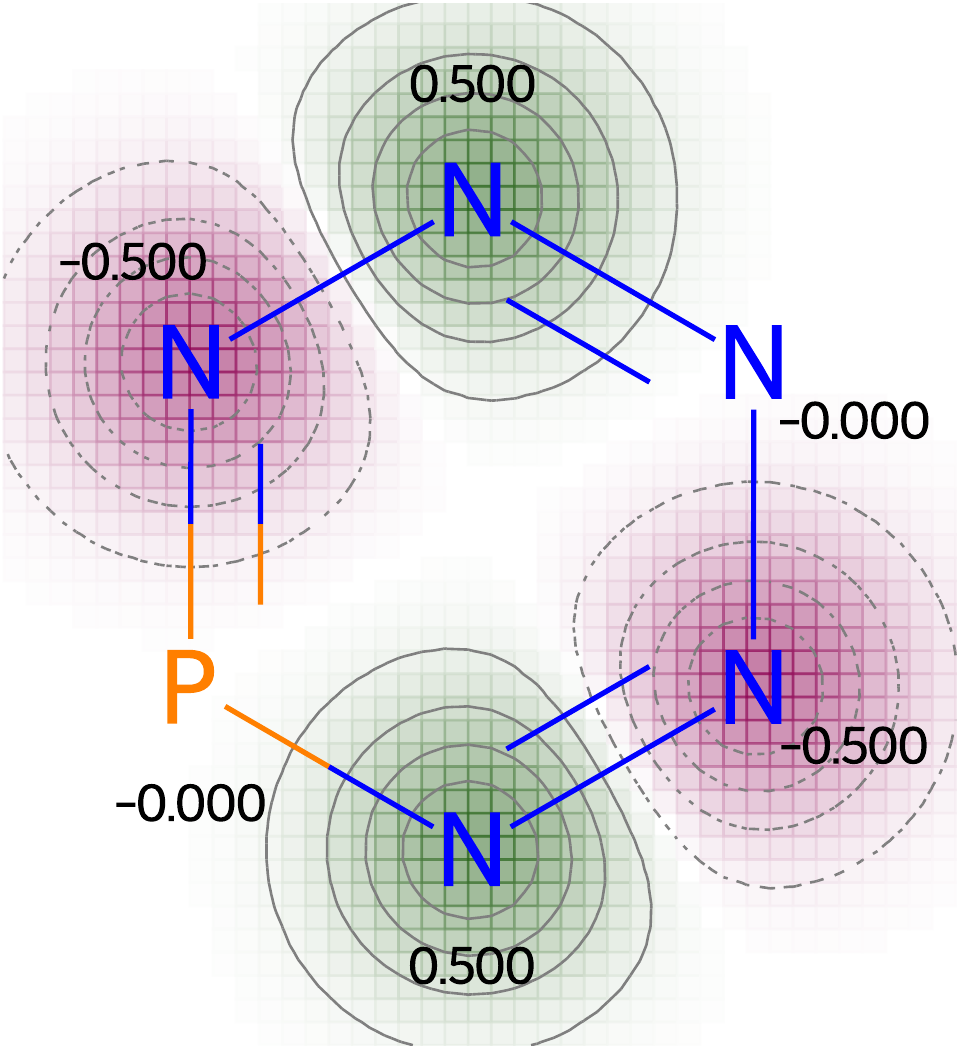}
    \end{subfigure}    
    {\small\textsf{\textbf{3}}}    
    \begin{subfigure}[a]{0.48\textwidth}
        \centering
        \includegraphics[height=3cm]{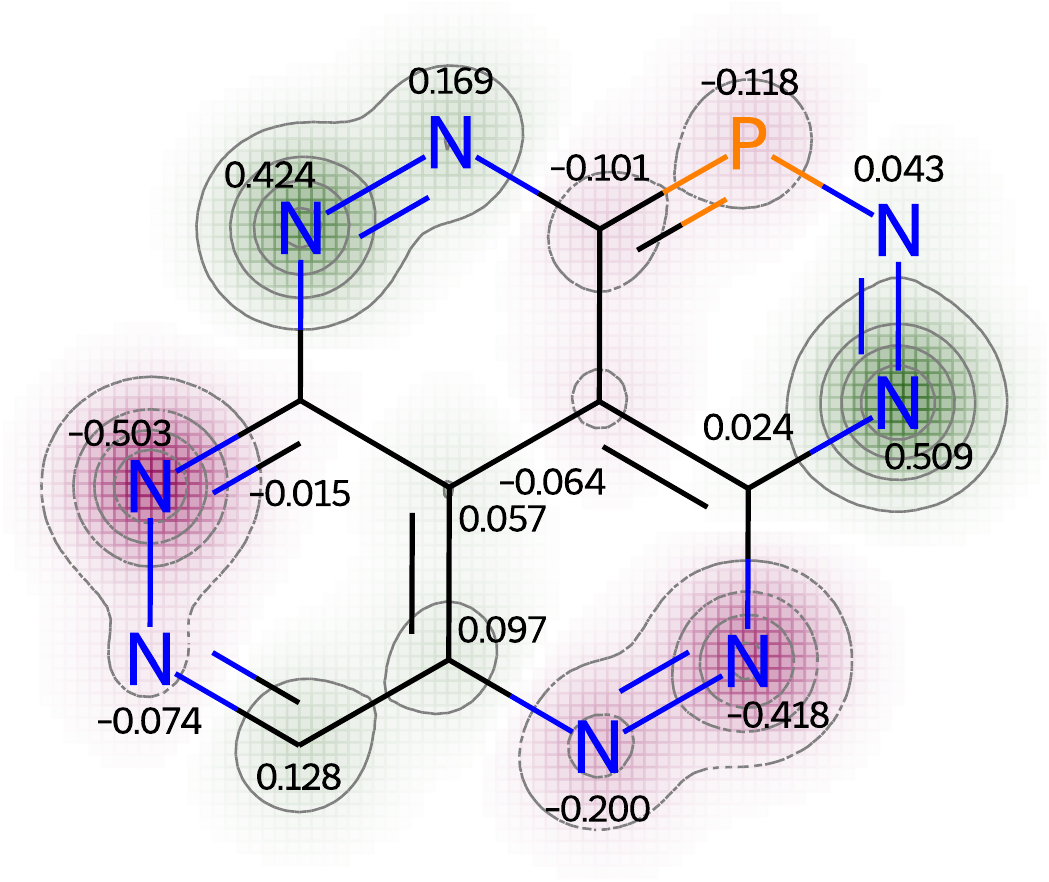}
    \end{subfigure}
    \begin{subfigure}[a]{0.48\textwidth}
        \centering
        \includegraphics[height=3cm]{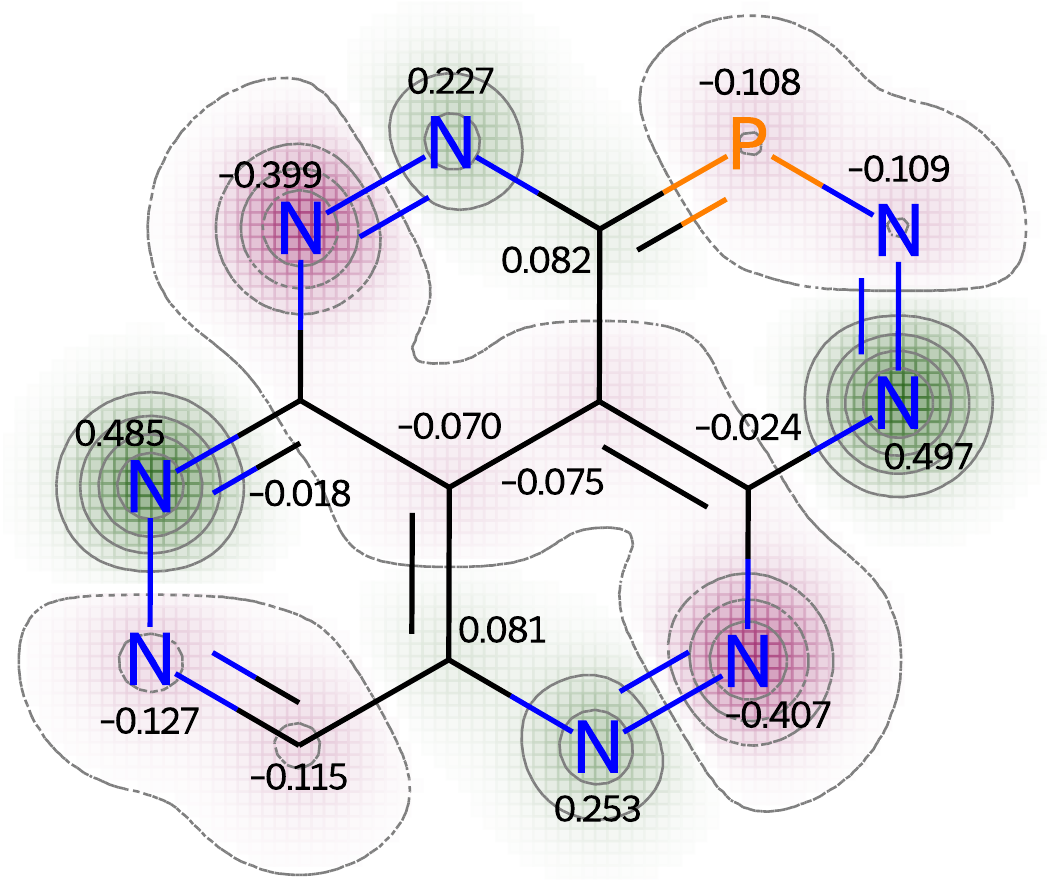}
    \end{subfigure}        
    {\small\textsf{\textbf{4}}}    
    \vspace{0.2cm}
    \begin{subfigure}[a]{0.48\textwidth}
        \centering
        \includegraphics[height=2cm]{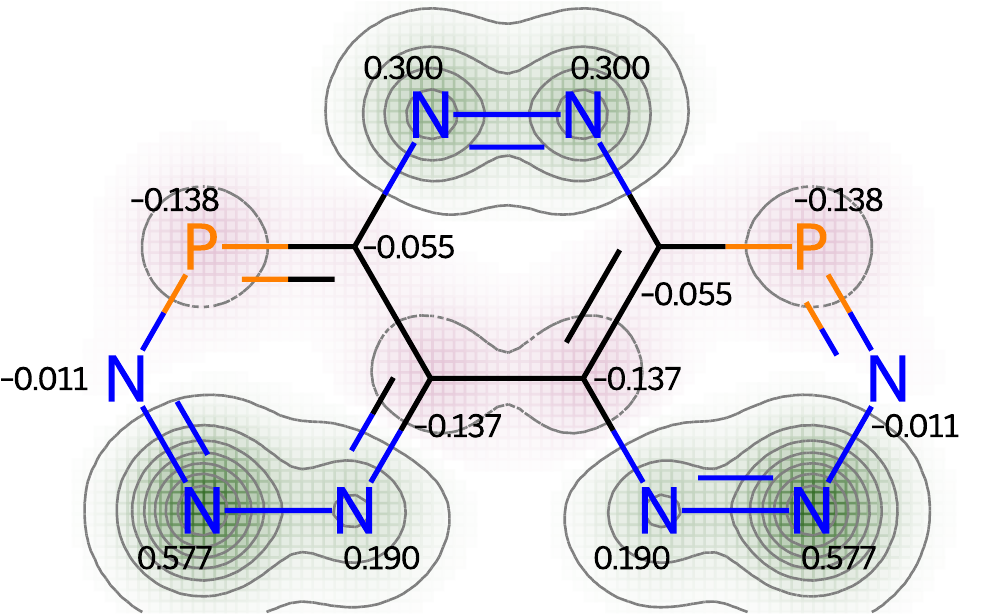}
    \end{subfigure}
    \begin{subfigure}[a]{0.48\textwidth}
        \centering
        \includegraphics[height=2cm]{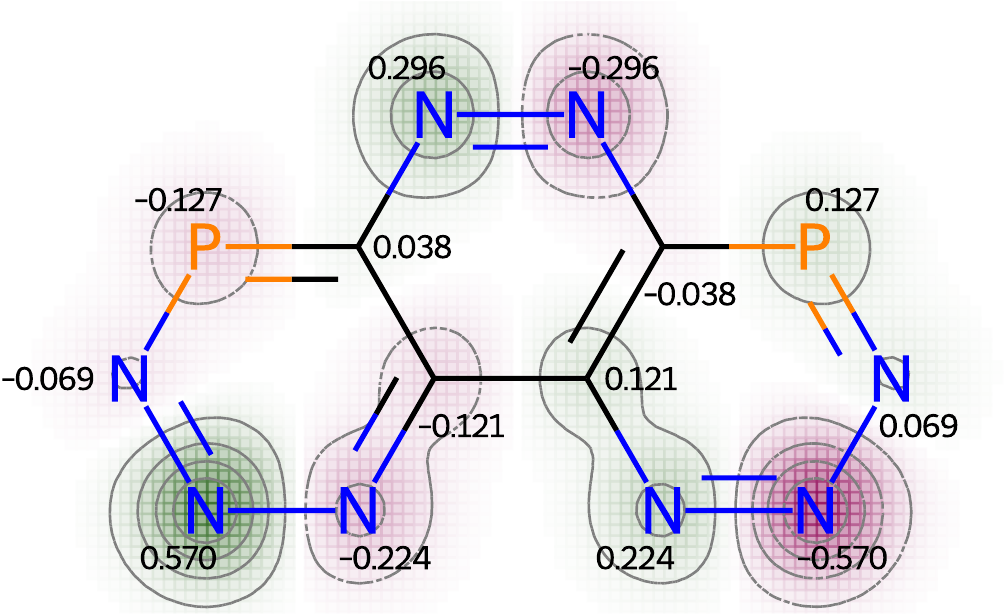}
    \end{subfigure}            
    {\small\textsf{\textbf{5}}}    
    \begin{subfigure}[a]{0.48\textwidth}
        \centering
        \includegraphics[height=1.7cm]{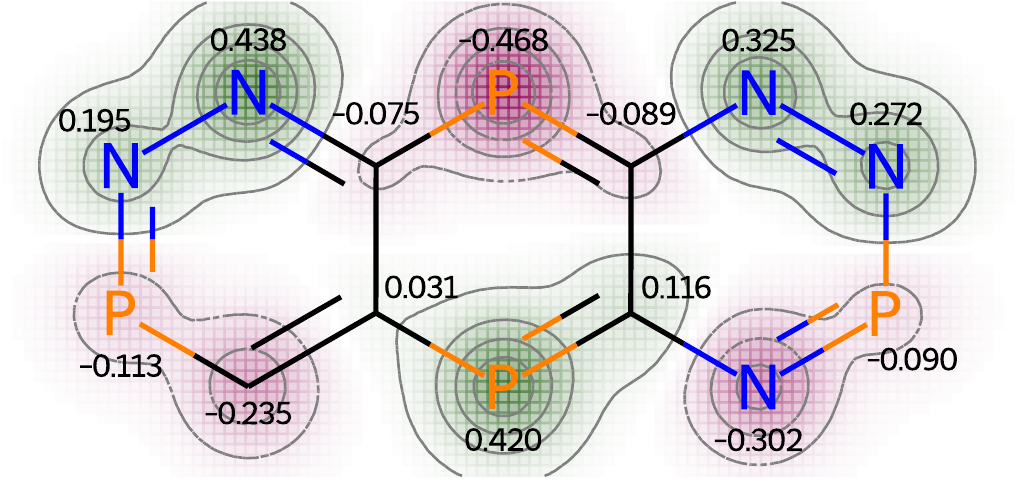}
    \end{subfigure}
    \begin{subfigure}[a]{0.48\textwidth}
        \centering
        \includegraphics[height=1.7cm]{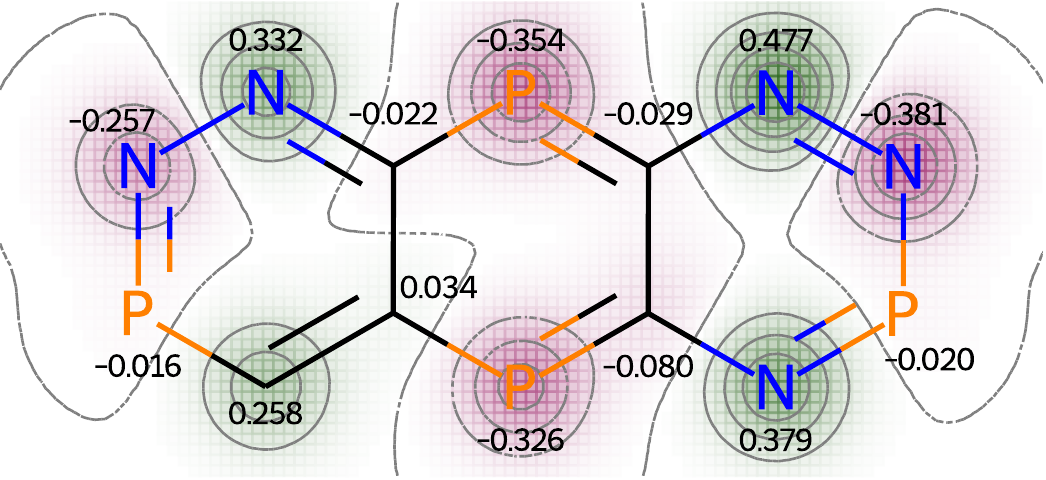}
    \end{subfigure}                
    {\small\textsf{\textbf{6}}}    
    \begin{subfigure}[a]{0.48\textwidth}
        \centering
        \includegraphics[height=3.5cm]{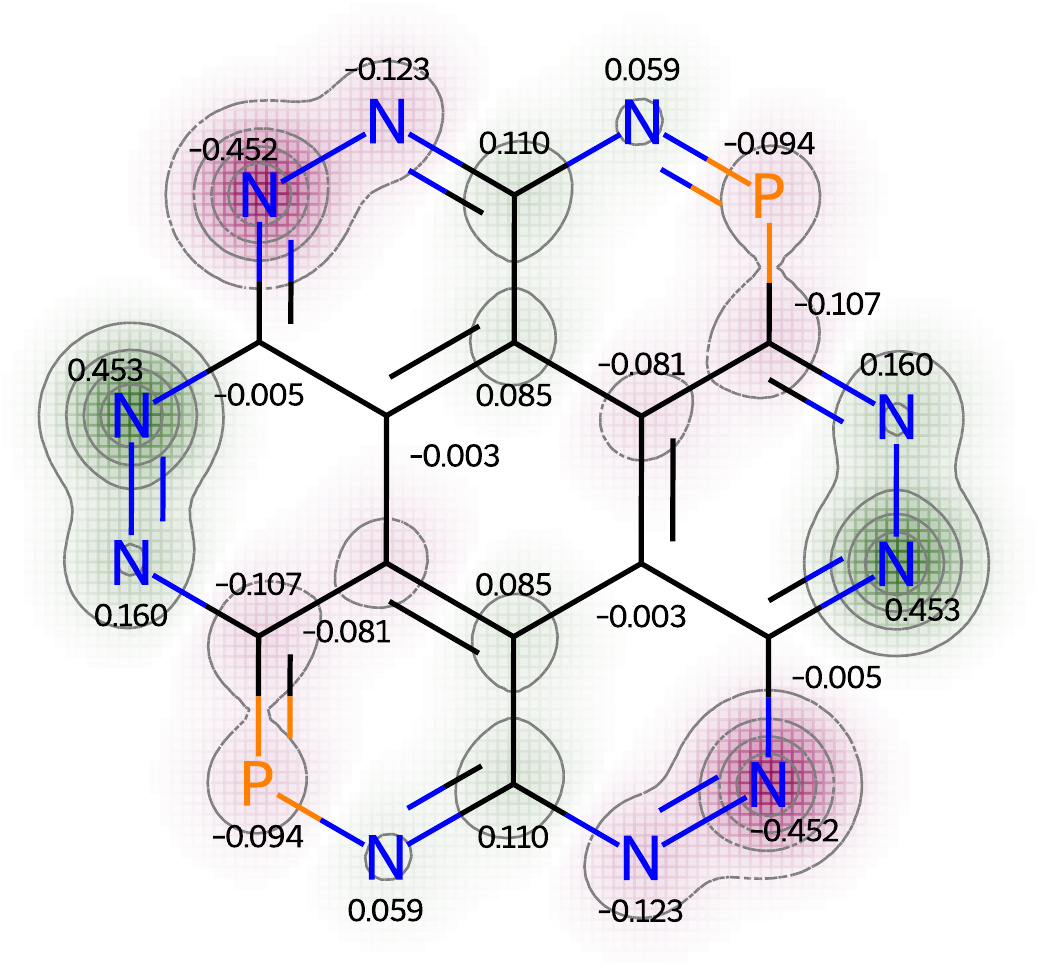}
    \end{subfigure}
    \begin{subfigure}[a]{0.48\textwidth}
        \centering
        \includegraphics[height=3.5cm]{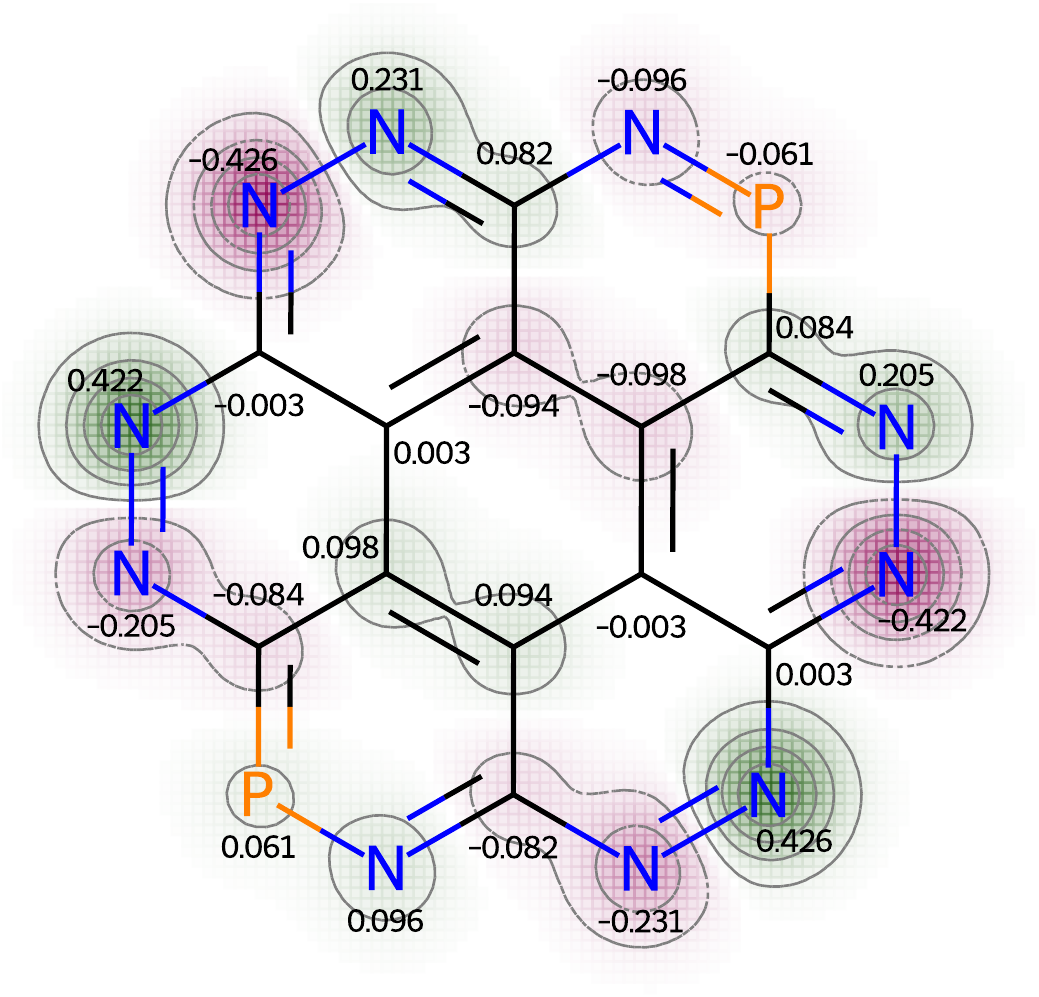}
    \end{subfigure}                    
    {\small\textsf{\textbf{7}}}    
    \begin{subfigure}[a]{0.48\textwidth}
        \centering
        \includegraphics[height=2.5cm]{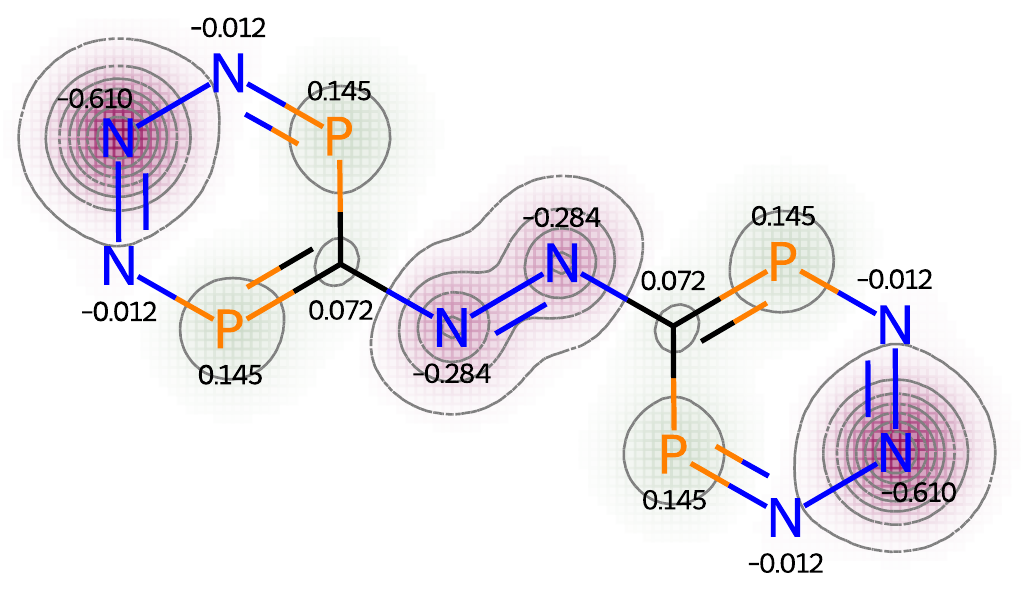}
    \end{subfigure}
    \begin{subfigure}[a]{0.48\textwidth}
        \centering
        \includegraphics[height=2.5cm]{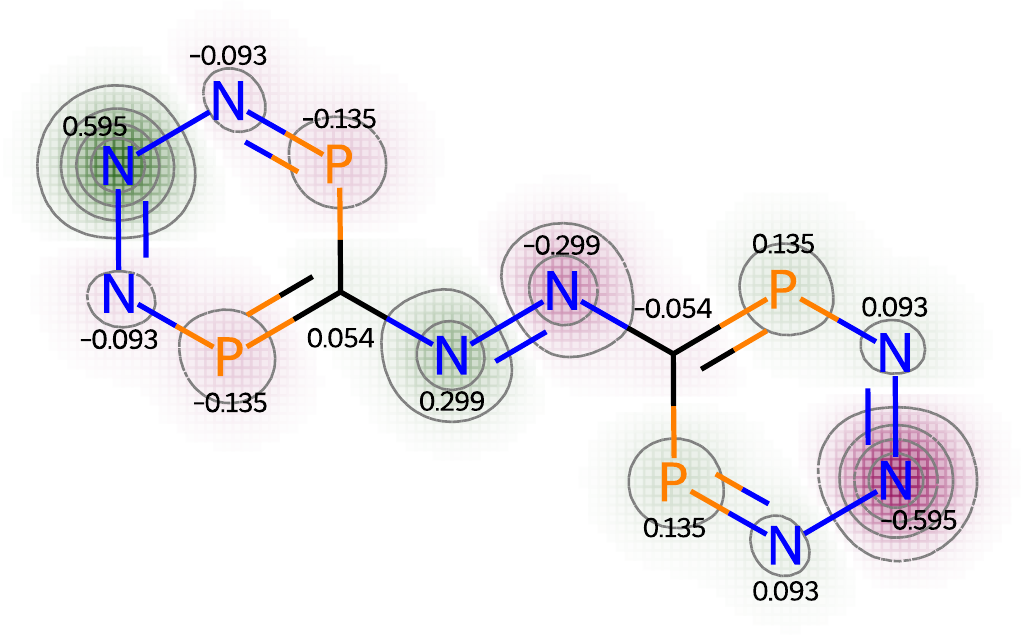}
    \end{subfigure}                        
    {\small\textsf{\textbf{8}}}    
    \begin{subfigure}[a]{0.48\textwidth}
        \centering
        \includegraphics[height=1.7cm]{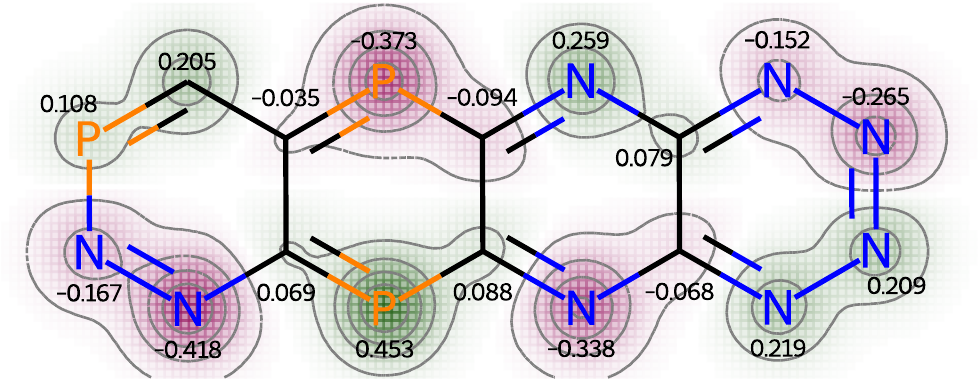}
    \end{subfigure}
    \begin{subfigure}[a]{0.48\textwidth}
        \centering
        \includegraphics[height=1.7cm]{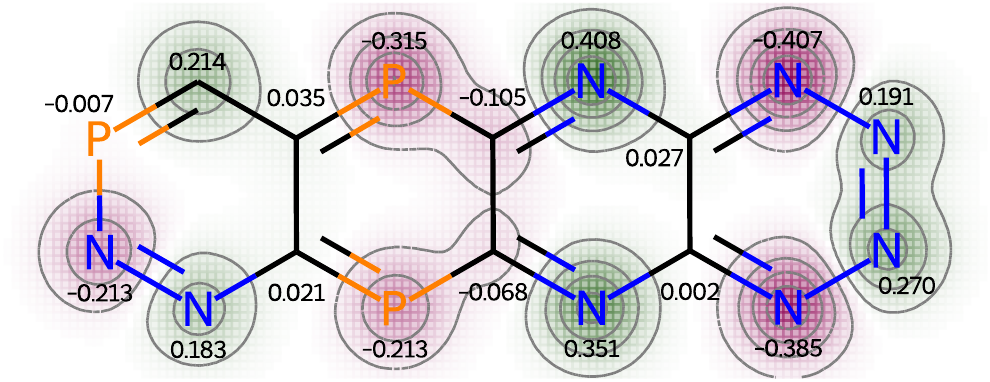}
    \end{subfigure}                            
    \caption{HOMO and LUMO depictions with coefficients for the best optimized molecules using the parameter set optimized for $\epsilon_{HL}$.}
    \label{fig:s_homo_lumo_opt}
\end{figure}

\begin{figure}
    \centering
    \begin{subfigure}[a]{0.49\textwidth}
        \centering
        \textsf{HOMO}
    \end{subfigure}
    \begin{subfigure}[a]{0.49\textwidth}
        \centering
        \textsf{LUMO}
    \end{subfigure}
    \vspace{0.5cm}        
    
    {\small\textsf{\textbf{1}}}
    \begin{subfigure}[a]{0.48\textwidth}
        \centering
        \includegraphics[height=0.9cm]{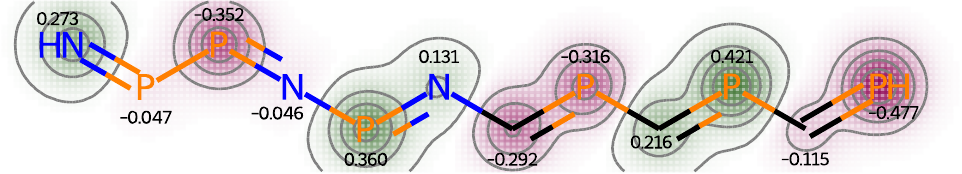}
    \end{subfigure}
    \begin{subfigure}[a]{0.48\textwidth}
        \centering
        \includegraphics[height=0.9cm]{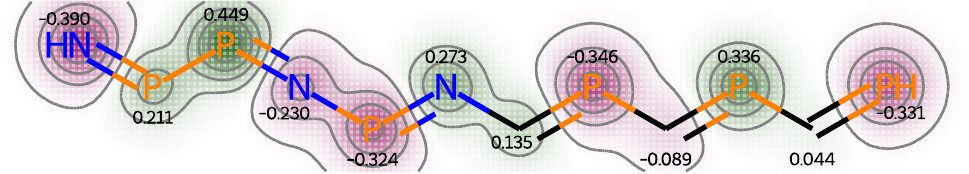}
    \end{subfigure}
    {\small\textsf{\textbf{2}}}
    \begin{subfigure}[a]{0.48\textwidth}
        \centering
        \includegraphics[height=2cm]{Figures/Figures_HL/Opt/2_HOMO.pdf}
    \end{subfigure}
    \begin{subfigure}[a]{0.48\textwidth}
        \centering
        \includegraphics[height=2cm]{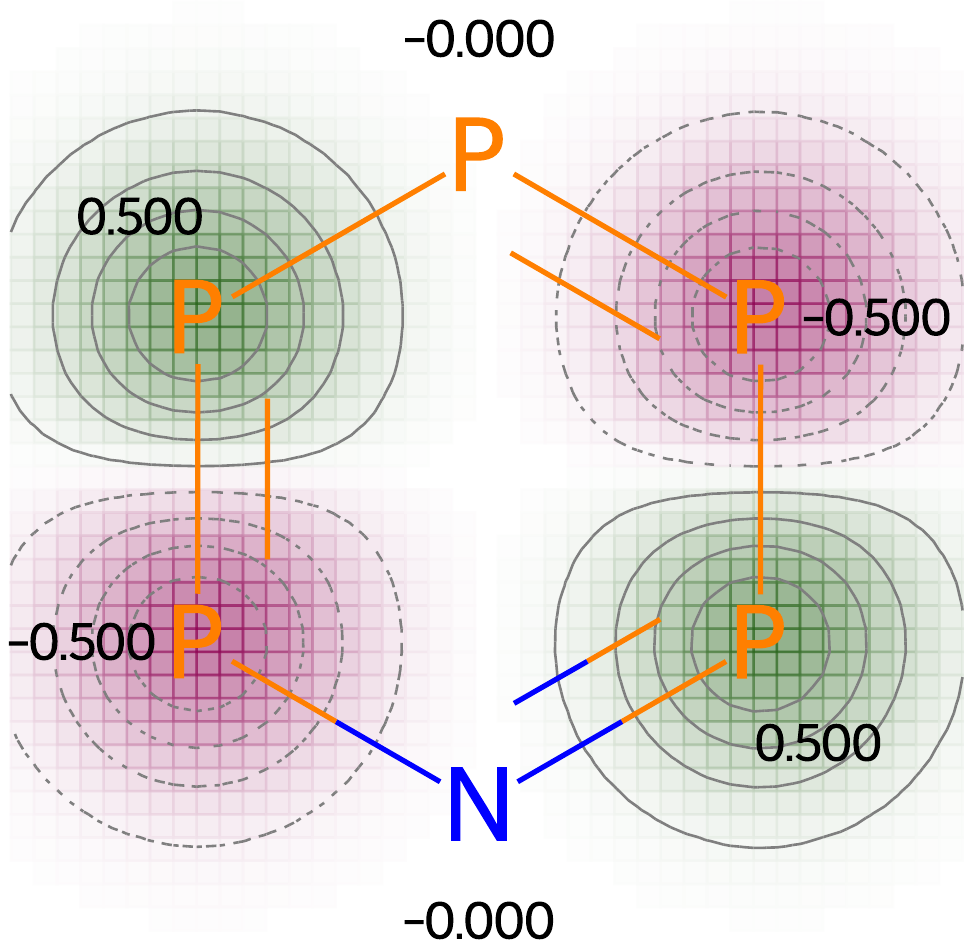}
    \end{subfigure}    
    {\small\textsf{\textbf{3}}}    
    \begin{subfigure}[a]{0.48\textwidth}
        \centering
        \includegraphics[height=3cm]{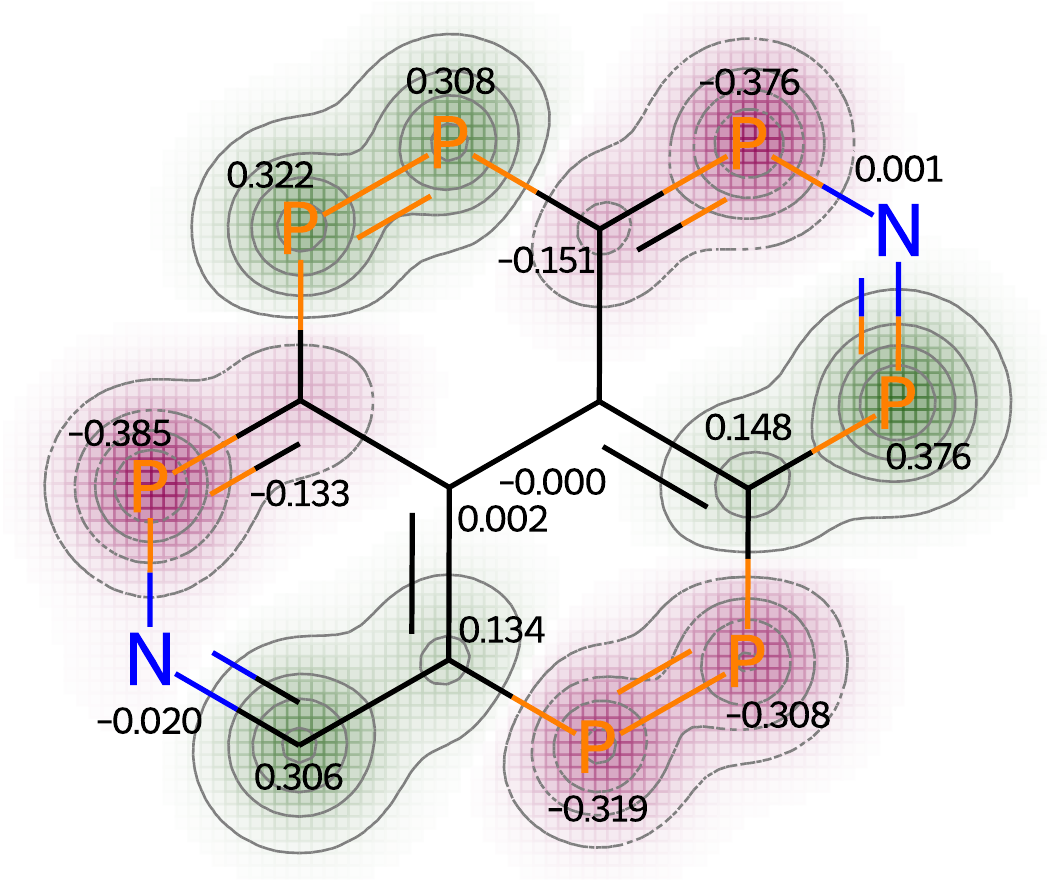}
    \end{subfigure}
    \begin{subfigure}[a]{0.48\textwidth}
        \centering
        \includegraphics[height=3cm]{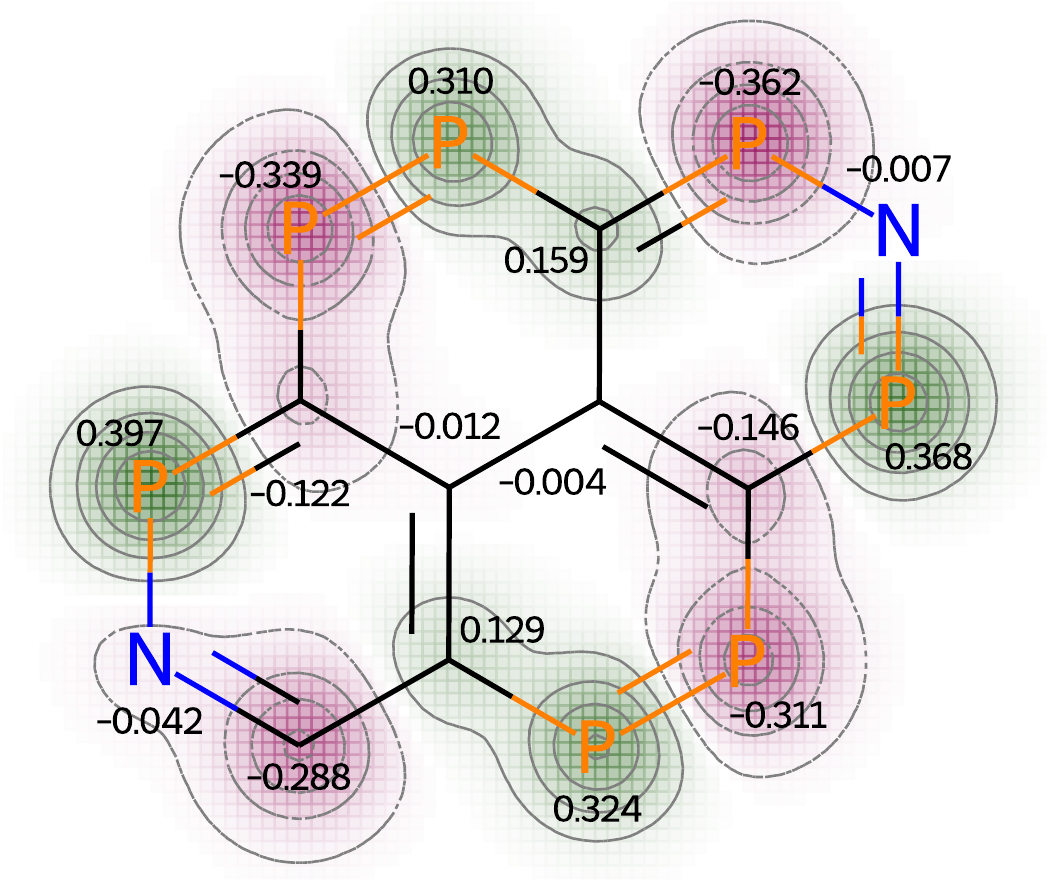}
    \end{subfigure}        
    {\small\textsf{\textbf{4}}}    
    \vspace{0.2cm}
    \begin{subfigure}[a]{0.48\textwidth}
        \centering
        \includegraphics[height=2cm]{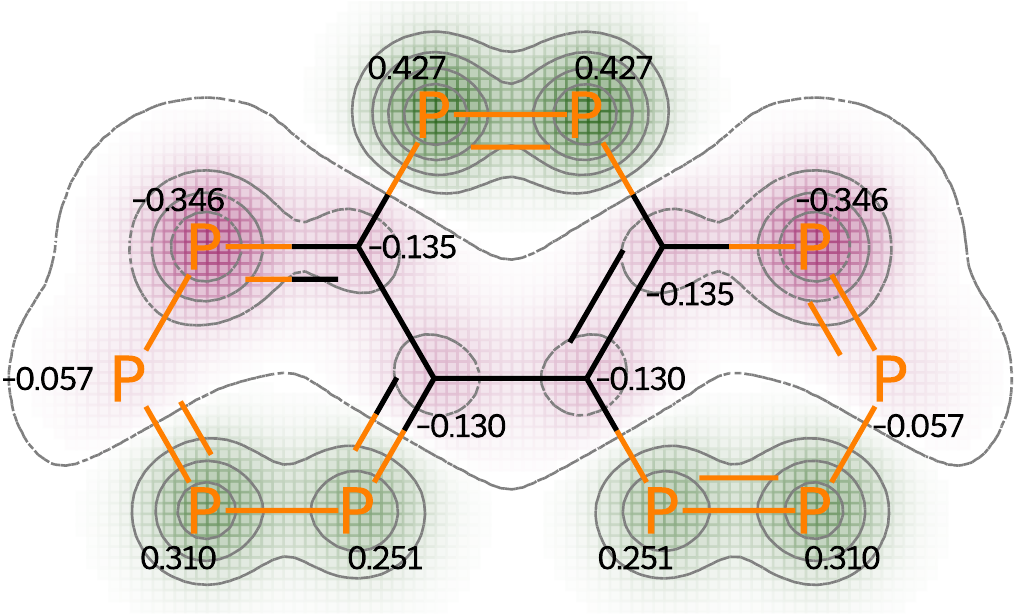}
    \end{subfigure}
    \begin{subfigure}[a]{0.48\textwidth}
        \centering
        \includegraphics[height=2cm]{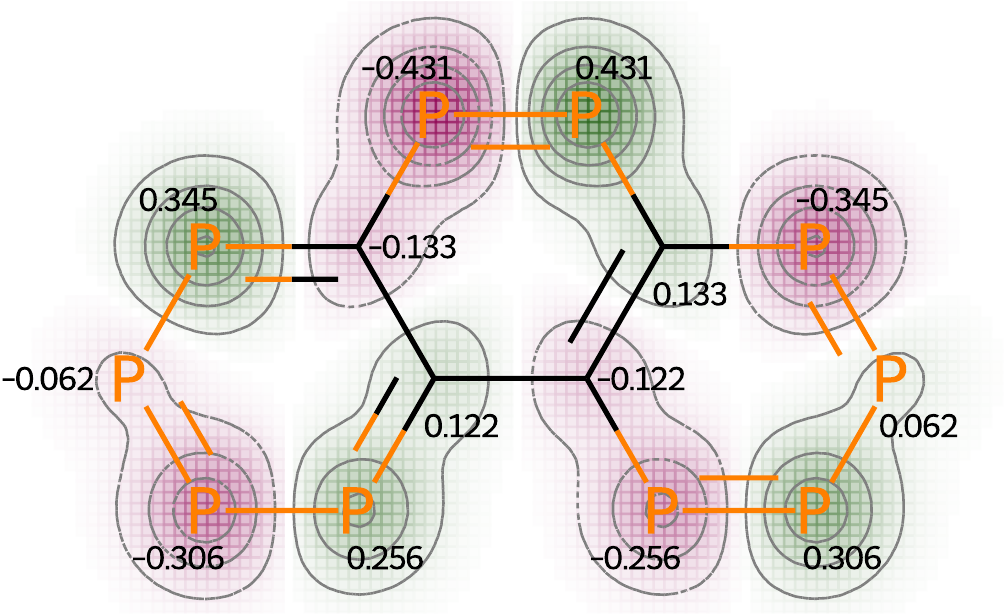}
    \end{subfigure}            
    {\small\textsf{\textbf{5}}}    
    \begin{subfigure}[a]{0.48\textwidth}
        \centering
        \includegraphics[height=1.7cm]{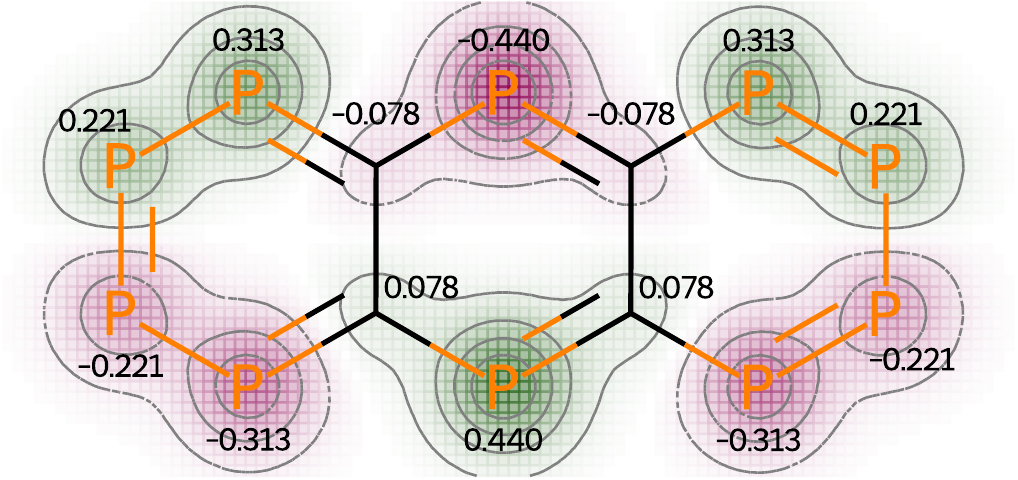}
    \end{subfigure}
    \begin{subfigure}[a]{0.48\textwidth}
        \centering
        \includegraphics[height=1.7cm]{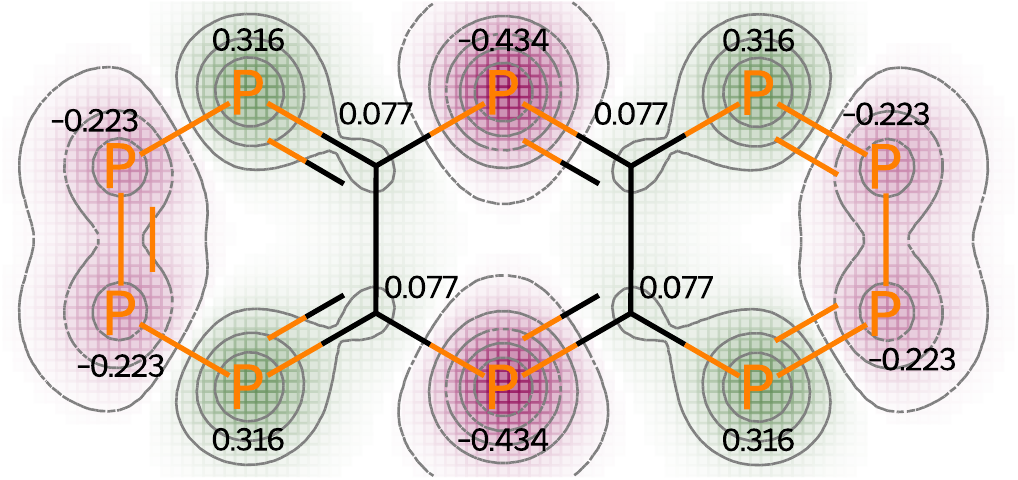}
    \end{subfigure}                
    {\small\textsf{\textbf{6}}}    
    \begin{subfigure}[a]{0.48\textwidth}
        \centering
        \includegraphics[height=3.5cm]{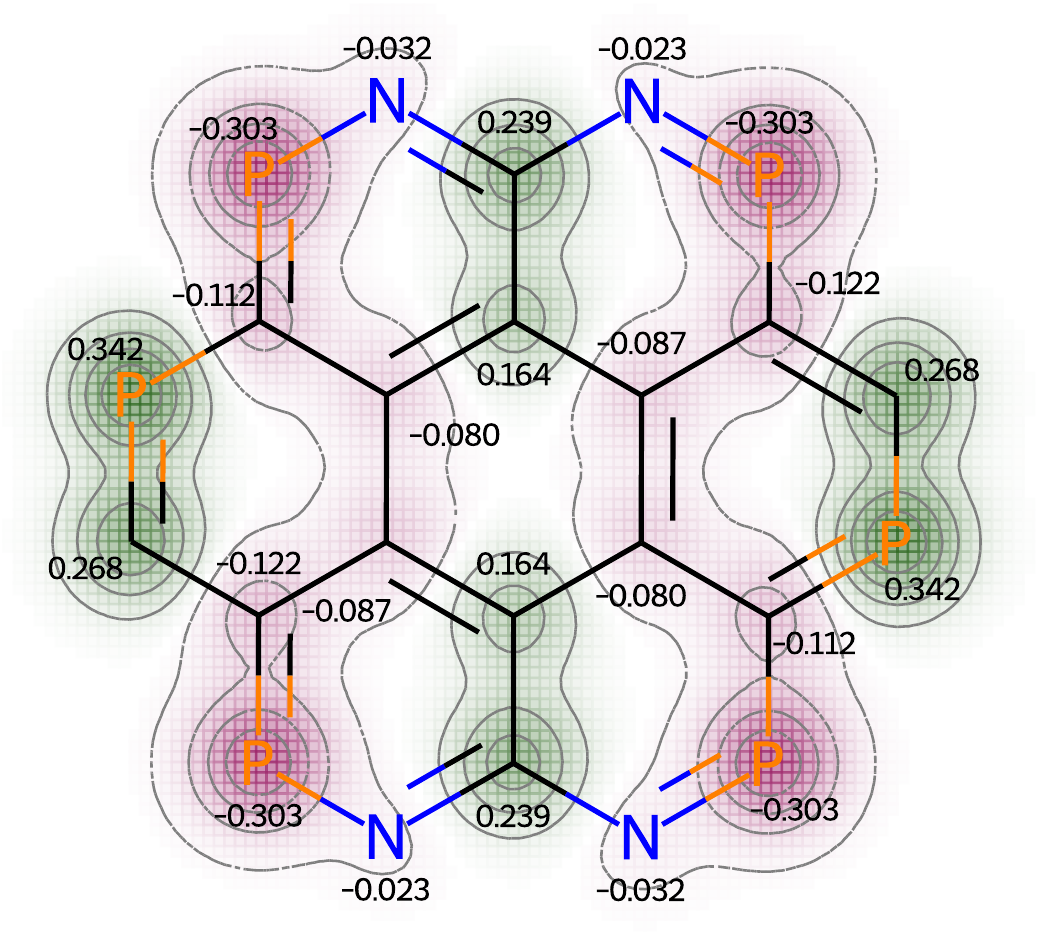}
    \end{subfigure}
    \begin{subfigure}[a]{0.48\textwidth}
        \centering
        \includegraphics[height=3.5cm]{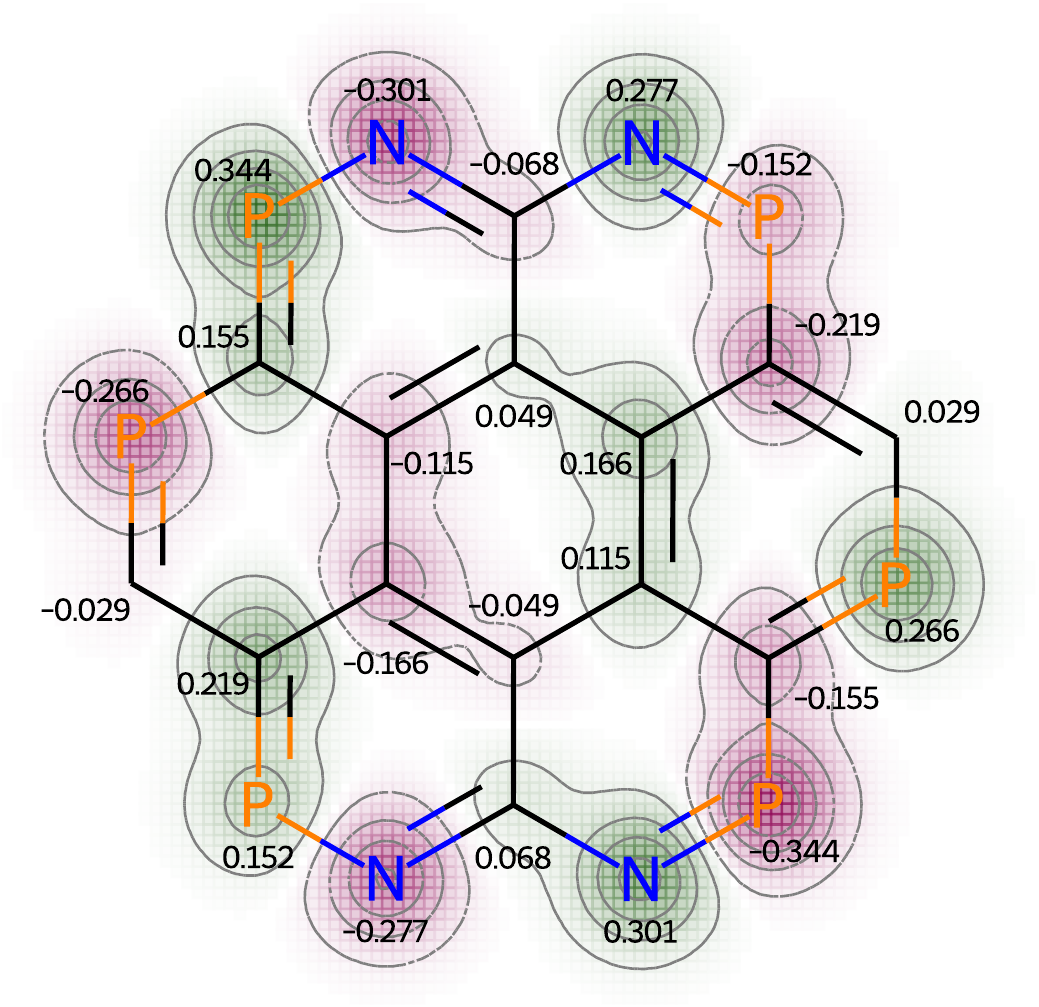}
    \end{subfigure}                    
    {\small\textsf{\textbf{7}}}    
    \begin{subfigure}[a]{0.48\textwidth}
        \centering
        \includegraphics[height=2.5cm]{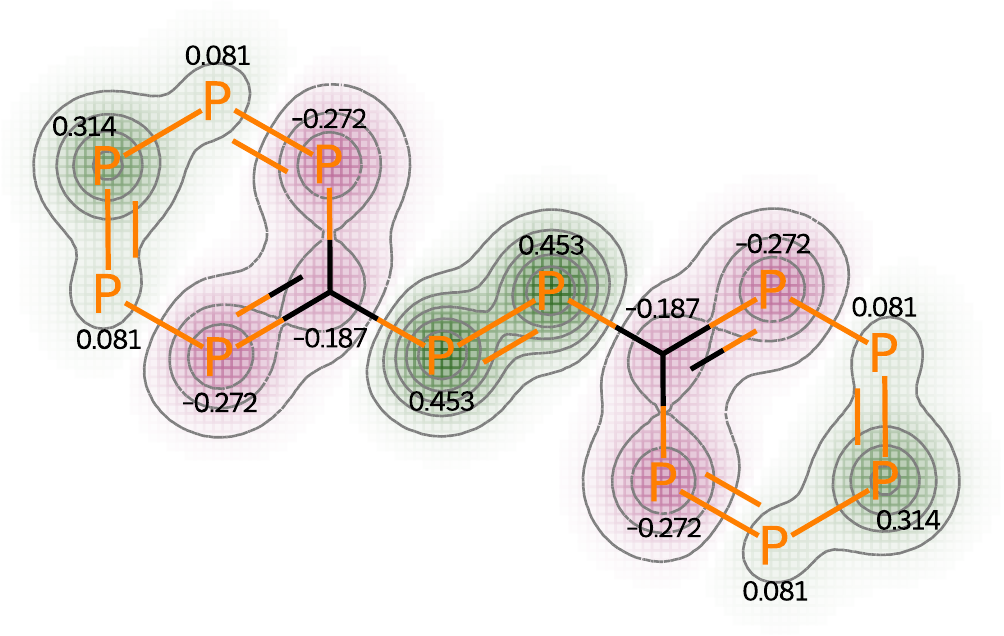}
    \end{subfigure}
    \begin{subfigure}[a]{0.48\textwidth}
        \centering
        \includegraphics[height=2.5cm]{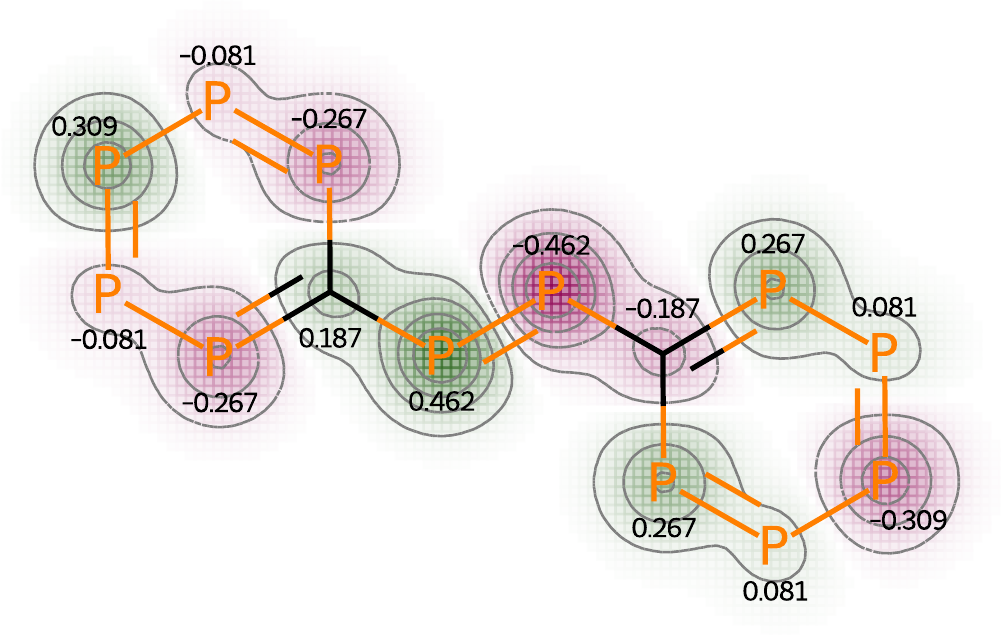}
    \end{subfigure}                        
    {\small\textsf{\textbf{8}}}    
    \begin{subfigure}[a]{0.48\textwidth}
        \centering
        \includegraphics[height=1.7cm]{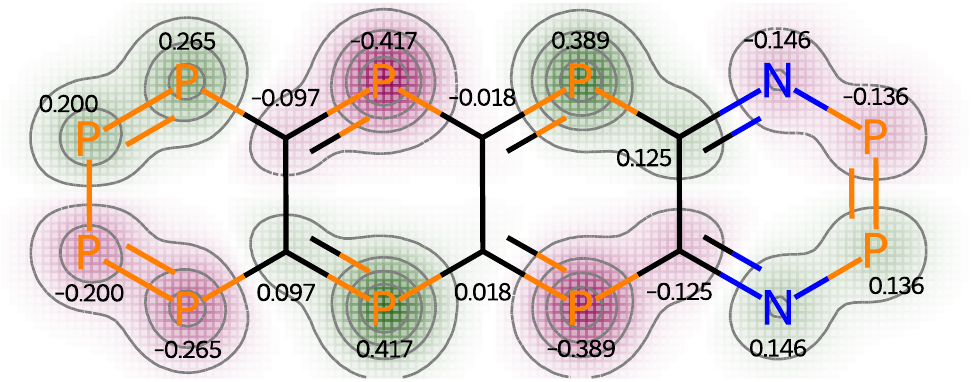}
    \end{subfigure}
    \begin{subfigure}[a]{0.48\textwidth}
        \centering
        \includegraphics[height=1.7cm]{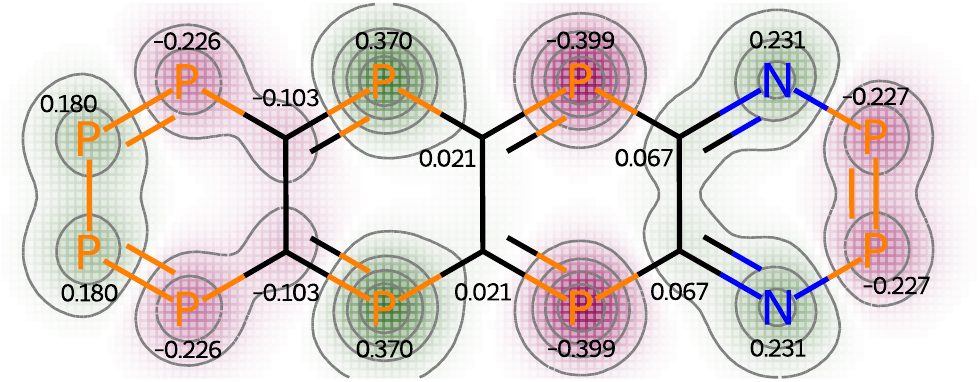}
    \end{subfigure}                            
    \caption{HOMO and LUMO depictions with coefficients for the best optimized molecules using the Van-Catledge parameter set.}
    \label{fig:s_homo_lumo_orig}
\end{figure}

\pagebreak

\section{References}
\bibliography{references} 